\documentclass[11pt]{article}              

\usepackage[margin=25mm]{geometry}
\usepackage{graphicx}
\usepackage{colortbl}
\usepackage[FIGBOTCAP]{subfigure}
 
\usepackage{fancyhdr}
\usepackage{amssymb,amsfonts,amsmath,amsthm}

\usepackage{authblk}
	
\graphicspath{{./},{./Figures/}}

\usepackage{bbm}
\usepackage{sudretstyle}
\usepackage{booktabs} 
\usepackage{multirow} 

\usepackage{color}

\usepackage{lineno}

\begin{document}
\title{Seismic fragility curves for structures using non-parametric representations} 

\author[1]{C. Mai} \author[1]{K. Konakli} \author[1]{B. Sudret} 

\affil[1]{Chair of Risk, Safety and Uncertainty Quantification,
  
  ETH Zurich, Stefano-Franscini-Platz 5, 8093 Zurich, Switzerland}

\date{}
\maketitle

\abstract{Fragility curves are commonly used in civil engineering to assess the vulnerability of structures to earthquakes. The probability of failure associated with a prescribed criterion (e.g. the maximal inter-storey 
	drift of a building exceeding a certain threshold) is represented as a function of the intensity of the earthquake ground motion (e.g. peak ground acceleration or spectral acceleration). The classical approach relies on assuming a lognormal shape of the fragility curves; it is thus parametric. In this paper, we introduce two non-parametric approaches to establish the fragility curves without employing the above assumption, namely binned Monte Carlo simulation and kernel density estimation. As an illustration, we compute the fragility curves for a three-storey steel frame using a large number of synthetic ground motions. The curves obtained with the non-parametric approaches are compared with respective curves based on the lognormal assumption. A similar comparison is presented for a case when a limited number of recorded ground motions is available. It is found that the accuracy of the lognormal curves depends on the ground motion intensity measure, the failure criterion and most importantly, on the employed method for estimating the parameters of the lognormal shape. \\[1em] 

}

\maketitle

\section{Introduction}
\label{intro}
The severe socio-economic consequences of several recent earthquakes highlight the need for proper seismic risk assessment as a basis for efficient decision making on mitigation actions and disaster planning. To this end, the probabilistic performance-based earthquake engineering (PBEE) framework has been developed, which allows explicit evaluation of performance measures that serve as decision variables (DV) (\eg monetary losses, casualties, downtime) accounting for the prevailing uncertainties (\eg ground motion characteristics, structural properties, damage occurrence). The key steps in the PBEE framework comprise the identification of seismic hazard, the evaluation of structural response, damage analysis and eventually, consequence evaluation. In particular, the mean annual frequency of exceedance of a DV is evaluated as \cite{Porter2003,Baker2008a,Gunay2013}:
\begin{equation}
	\lambda(DV)=\int\int\int P(DV|DM) \, \di P(DM|EDP) \, \di P(EDP|IM) \, \abs{\di \lambda(IM)},
	\label{eqPBEE}
\end{equation}
in which $P(x|y)$  is the conditional probability of $x$ given $y$, $DM$  is a damage measure typically defined according to repair costs (\eg light, moderate or severe damage), $EDP$ is an engineering demand parameter obtained from structural analysis (\eg force, displacement, drift ratio), $IM$ is an intensity measure characterizing the ground motion severity (\eg peak ground acceleration, spectral acceleration) and $\lambda(IM)$ is the annual frequency of exceedance of the $IM$. Determination of the probabilistic model $P(EDP|IM)$ constitutes a major challenge in the PBEE framework since the earthquake excitation contributes the most significant part to the uncertainty in the $DV$. The present paper is concerned with this step of the analysis.

The conditional probability $P(EDP \geq \overline{edp}|IM)$, where $\overline{edp}$ denotes an acceptable demand threshold, is commonly represented graphically in the shape of the so-called demand fragility curves \cite{Mackie2005}. Thus, a demand fragility curve represents the probability that an engineering demand parameter exceeds a prescribed threshold as a function of an intensity measure of the earthquake motion. For the sake of simplicity, demand fragility curves are simply denoted fragility curves hereafter, which is also typical in the literature \cite{Ellingwood2009,Seo2012}. We note however that the term \emph{fragility} may also be used for $P(DM\geq\overline{dm}|IM)$ or $P(DM\geq\overline{dm}|EDP)$, \ie the conditional probability of the damage measure exceeding a threshold $\overline{dm}$ given the ground motion intensity \cite{Banerjee2007} or the engineering demand parameter \cite{Baker2008a,Gunay2013}, respectively.

Originally introduced in the early 1980's for nuclear safety evaluation \cite{Richardson1980}, fragility curves are nowadays widely used for multiple purposes, \eg loss estimation \cite{Pei2009}, assessment 
of collapse risk \cite{Eads2013}, design checking \cite{Dukes2012}, evaluation of the effectiveness of retrofit measures \cite{Guneyisi2008}, etc. Several novel methodological contributions to fragility analysis have been made in recent years, including the development of multi-variate fragility functions \cite{Seyedi2010}, the incorporation of Bayesian updating \cite{Gardoni2002a} and the consideration of time-dependent fragility \cite{Ghosh2010}. However, the traditional fragility curves remain a popular tool in seismic risk assessment and recent literature is rich with relevant applications on various type of structures, such as irregular buildings \cite{Seo2012}, underground tunnels \cite{Argyroudis2012}, pile-supported wharfs \cite{Chiou2011}, wind turbines \cite{Quilligan2012}, nuclear power plant equipments \cite{Borgonovo2013}, masonry buildings \cite{Karantoni2014}. The estimation of such curves is the focus of the present paper.

Fragility curves are typically classified into four categories according to the data sources, namely analytical, empirical, judgment-based or hybrid fragility curves \cite{Rossetto2005}. Analytical fragility curves are derived from data obtained by analyses of structural models. Empirical fragility curves are based on the observation of earthquake-induced damage reported in post-earthquake surveys. Judgment-based curves are estimated by expert panels specialized in the field of earthquake engineering. Hybrid curves are obtained by combining data from different sources. Each of the aforementioned categories has its own advantages and limitations. In this paper, analytical fragility curves based on data collected from numerical structural analyses are of interest.

The typical approach to compute analytical fragility curves presumes that the curves have the shape of a lognormal cumulative distribution function \cite{Shinozuka2000b,Ellingwood2001}. This approach is therefore considered \emph{parametric}. The parameters of the lognormal distribution are determined either by maximum likelihood 
estimation  \cite{Shinozuka2000b,Zentner2010a,Seyedi2010} or by fitting a linear probabilistic seismic demand model in the log-scale \cite{Ellingwood2009,Gencturk2008,Jeong2012,Banerjee2008}.
The assumption of lognormal fragility curves is almost unanimous in the literature due to the computational convenience as well as due to the ease of combining such curves with other elements of the seismic probabilistic risk assessment framework. However, the validity of such assumption remains questionable (see also \cite{Karamlou2015}).

In this paper, we present two non-parametric approaches for establishing the fragility curves, namely binned Monte Carlo simulation (bMCS) and kernel density estimation (KDE). The main advantage of bMCS over existing techniques also based on Monte Carlo simulation is that it avoids the bias induced by scaling ground motions to predefined intensity levels. In the KDE approach, we introduce a statistical methodology for fragility estimation, which opens new paths for estimating multi-dimensional fragility functions as well. The proposed methods are subsequently used to investigate the validity of the lognormal assumption in a case study, where we develop fragility curves for different thresholds of the maximum drift ratio of a three-story steel frame subject to synthetic ground motions. The comparison between KDE-based and lognormal fragility curves is also shown for a concrete bridge column subject to recorded motions using results from an earlier study by the authors \cite{MaiEurodyn2014}. The proposed methodology can be applied in a straightforward manner to other types of structures or classes of structures or using different failure criteria.

The paper is organized as follows: in Section 2, the different approaches for establishing the fragility curves, namely the lognormal and the proposed bMCS and KDE approaches, are presented. In Section 3, the method recently developed by Rezaeian and Der Kiureghian \cite{Rezaeian2008} for generating synthetic ground motions, which is employed in the following numerical investigations, is briefly recalled. The case studies are presented in Sections 4 and 5 and the results are discussed in Section 6. The paper concludes with a summary of the main findings and perspectives on future research.

\section{Computation of fragility curves}
\label{sec3}
Fragility curves represent the probability of failure of a system, associated with a specified criterion, for a given intensity measure ($IM$) of the earthquake motion. Failure herein represents the exceedence of a prescribed demand limit. A commonly used demand parameter in earthquake engineering is the maximal drift ratio $\Delta$, \ie the maximal relative horizontal displacement normalized by the corresponding height \cite{Seo2012}. Thus, the fragility function is cast as follows:
\begin{equation}
	\text{Frag}(IM;\,\delta_o)=\mathbb{P}[\Delta\geq \delta_o \lvert IM],
	\label{eq7}
\end{equation}
in which $\text{Frag}(IM;\,\delta_o)$ denotes the fragility at the given $IM$ for a threshold $\delta_o$ of $\Delta$. In order to establish the fragility curves, a number $N$ of transient finite element analyses of the structure under consideration are used to provide paired values $\acc{\prt{IM_i,\Delta_i}, i=1 \enu N}$.
\subsection{Lognormal approach}
\label{sec3.1}
The classical approach for establishing fragility curves consists in
assuming a \emph{lognormal shape} for the curves described in \eqrefe{eq7}. Two techniques are typically used to estimate the parameters of the lognormal fragility curves, namely maximum likelihood estimation and linear regression. These are presented below.

\subsubsection{Maximum likelihood estimation}

One assumes that the fragility curves can be written in the following general form:
\begin{equation}
	\widehat{\text{Frag}}(IM;\,\delta_o) = \Phi \prt{\dfrac{\ln IM - \ln \alpha}{\beta}},
	\label{eq9b}
\end{equation}
where $\Phi(\cdot)$ denotes the standard Gaussian
cumulative distribution function (CDF), $\alpha$ is the ``median'' and $\beta$ is the ``log-standard deviation'' of the lognormal curve.
Shinozuka et al. \cite{Shinozuka2000b} proposed the use of maximum likelihood estimation to determine these parameters as follows: One denotes by $\omega$ the event that the demand threshold $\delta_o$ is reached or exceeded and assumes that $Y(\omega)$ is a random variable with a Bernoulli distribution. In particular, $Y$ takes the value 1 with probability $\text{Frag}(\cdot;\,\delta_o)$ and the value 0 with probability $1-\text{Frag}(\cdot;\,\delta_o)$. Considering a set of $i=1 \enum N$ ground motions, the likelihood function reads:
\begin{equation}
	\cl \prt{\alpha,\,\beta,\,\acc{IM_i,\,i=1 \enum N}} = \prod_{i=1}^N \bra{\text{Frag}(IM_i;\,\delta_o)}^{y_i} \, \bra{1- \text{Frag}(IM_i;\,\delta_o)}^{1- y_i},
	\label{eq9c}
\end{equation}
where $IM_i$ is the intensity measure of the $i^{th}$ seismic motion and $y_i$ represents a realization of the Bernoulli random variable $Y$. The latter takes the value 1 or 0 depending on whether the structure under the $i^{th}$ ground motion sustains the demand threshold $\delta_o$ or not. The parameters $(\alpha,\, \beta)$ are obtained by maximizing the likelihood function. In practice, a straightforward optimization algorithm is applied on the log-likelihood function:
\begin{equation}
	\acc{\alpha^{\ast};\,\beta^{\ast}}\tr = \arg\max  \ln \cl \prt{\alpha,\,\beta,\,\acc{IM_i,\,i=1 \enum N}}.
	\label{eq9e}
\end{equation}

\subsubsection{Linear regression}
\label{sec2.1.2}

One first assumes a \textit{probabilistic seismic demand model}, which relates a structural response quantity of interest (herein drift ratio) to an intensity measure of the earthquake motion. Specifically, the demand $\Delta$ is assumed to follow a lognormal distribution of which the log-mean value is a \emph{linear function} of $\ln IM$, leading to:
\begin{equation}
	\ln \Delta = A \, \ln IM + B + \zeta \,Z,
	\label{eq8}
\end{equation}
where $Z \sim \cn(0,1)$ is a standard normal variable. Parameters $A$ and $B$ are determined by means of ordinary least squares estimation in a log-log scale. Parameter $\zeta$ is obtained by:
\begin{equation}
	\zeta^2= \sum_{i=1}^{N} e_i^2 /\prt{N-2},
\end{equation}
where $e_i$ the residual between the actual value $\ln \Delta$ and the value predicted by the linear model: $e_i= \ln \Delta_i - A \ln \prt{IM_i} - B$. Then, \eqrefe{eq7} rewrites:
\begin{equation}
	\begin{aligned}
		\widehat{\text{Frag}}(IM;\,\delta_o) &= \mathbb{P} \left[ \ln \Delta  \geq \ln \delta_o \right] 
		= 1- \mathbb{P} \left[ \ln \Delta  \leq \ln \delta_o \right]  \\
		& = \Phi \prt{\dfrac{\ln IM - \prt{\ln \delta_o - B}/A}{\zeta/A}}.
	\end{aligned}
	\label{eq9}
\end{equation}
A comparison to \eqrefe{eq9b} shows that the median and log-standard deviation of the lognormal fragility curve in \eqrefe{eq9} are $\alpha=\expo{\prt{\ln \delta_o - B}/A}$
and $\beta= \zeta/A$, respectively. This approach to fragility estimation is widely employed in the literature, see \eg \cite{Ellingwood2001,Choi2004,Padgett2008,Zareian2007} among others.

The two methods described in this section are \emph{parametric} because they impose the shape of the fragility curves (\eqrefe{eq9b} and \eqrefe{eq9}), which is that of a lognormal CDF when considered as a 
function of $IM$. We note that by using the linear-regression approach, one accepts two additional assumptions, namely the linear function for the log-mean value of $\Delta$ and the constant dispersion (or homoscedasticity) of the residuals independently of the $IM$ level. Effects of these assumptions have been investigated by Karamlou and Bocchini \cite{Karamlou2015}. In the sequel, we propose two \emph{non-parametric} approaches to compute fragility curves \emph{without} relying on the lognormality assumption.
\subsection{Binned Monte Carlo simulation}
\label{sec3.2}
Having at hand a large sample set $\acc{\prt{IM_j,\Delta_j},\,j=1 \enu N }$, it is possible to use \emph{binned Monte Carlo simulation} (bMCS) to compute the fragility curves, as described next. Let us consider a given abscissa $IM_o$. Within a small bin surrounding $IM_o$, say $\bra{IM_o -h, IM_o +h}$ one assumes that the maximal drift $\Delta$ is linearly related to the $IM$. This assumption is exact in the case of linear structures, but would only be an approximation in the nonlinear case. Therefore, the maximal drift $\Delta_j$, which is related to $IM_j \in \bra{IM_o -h, IM_o +h}$, is converted into the drift $\widetilde{\Delta_j}(IM_o)$, which is related to the $j^{th}$ input signal scaled to have an intensity measure equal to $IM_o$:
\begin{equation}
	\widetilde{\Delta_j}(IM_o)=\Delta_j \dfrac{IM_o}{IM_j}.
	\label{eq13}
\end{equation}
\begin{figure}[!ht]
	\centering
	\subfigure
	{
		\includegraphics[width=0.45\textwidth]{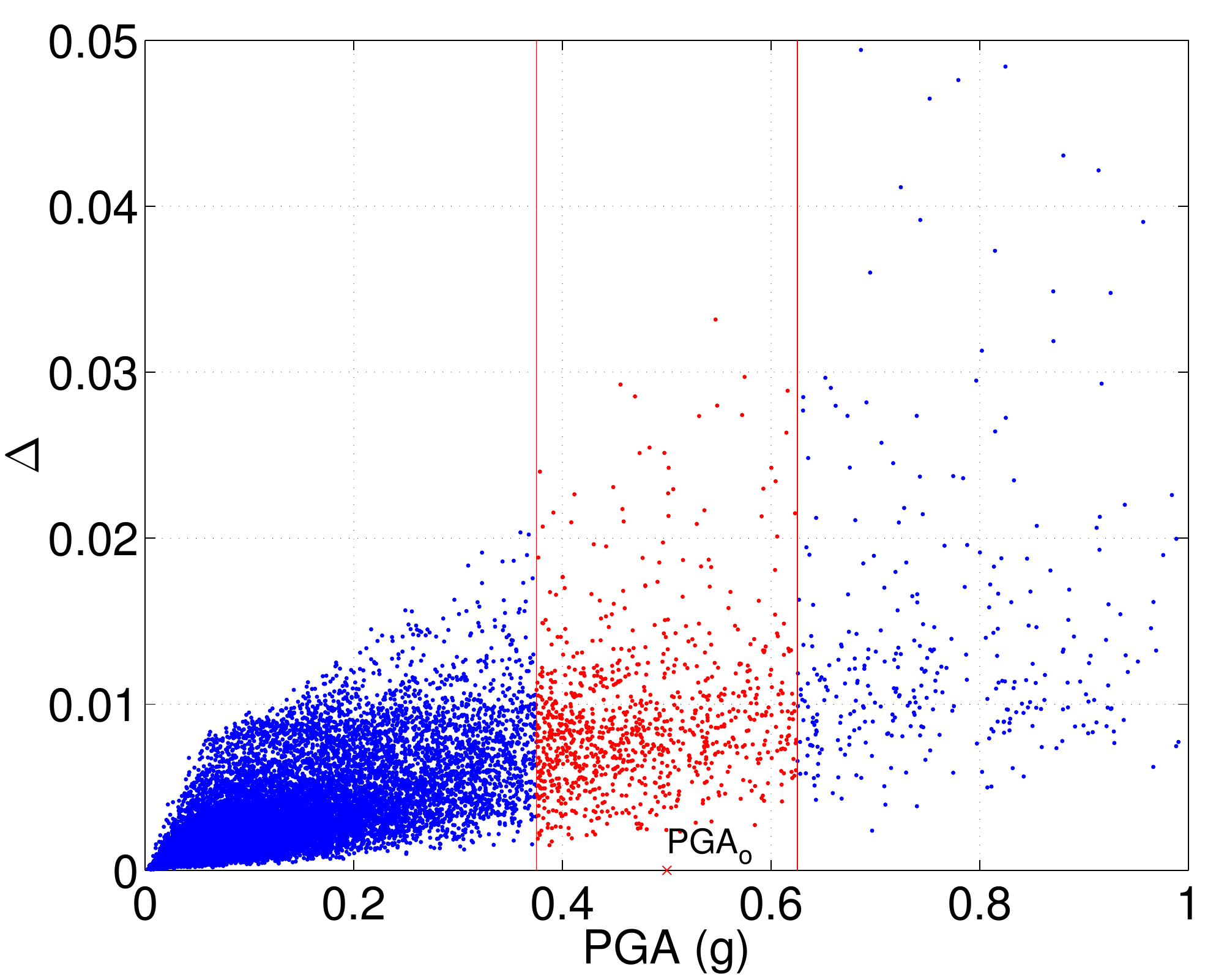}
	}
	\subfigure
	{
		\includegraphics[width=0.45\textwidth]{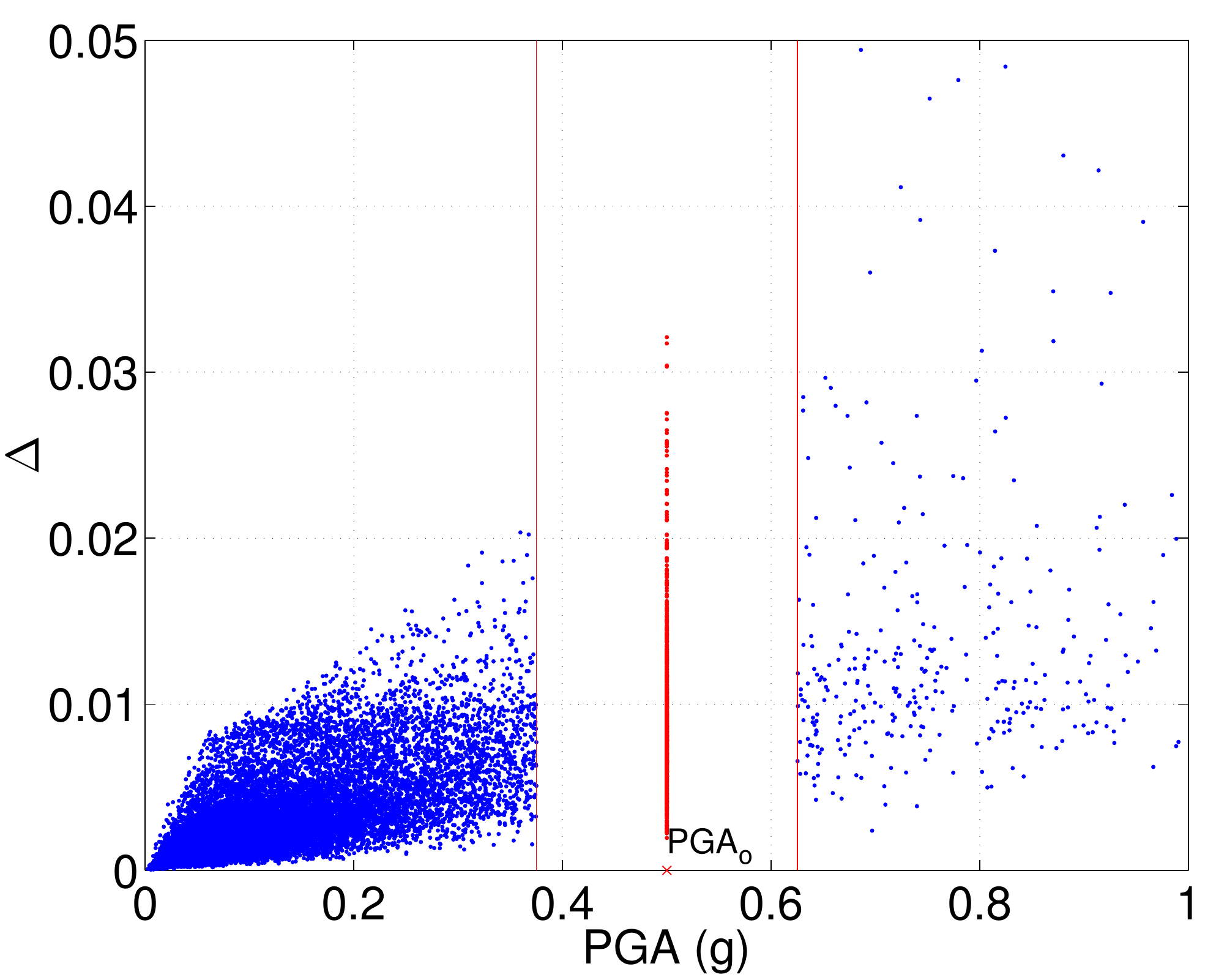}
	}
	\caption{(Left) Maximal drifts versus $IM$ before scaling.  (Right) Maximal drifts versus $IM$ after scaling within the bin marked with red color. (Note: a large bin is considered in the figure only to facilitate visualization of the method.)}
	\label{fig1}
\end{figure}
This procedure is illustrated in \figref{fig1}. The fragility curve at $IM_o$ is then obtained by a crude Monte Carlo estimator:
\begin{equation}
	\widehat{\text{Frag}} (IM_o)=\frac{N_f\prt{IM_o}}{N_s\prt{IM_o}},
\end{equation}
where $N_f\prt{IM_o}$ is the number of points in the bin such that $ \widetilde{\Delta_j}(IM_o) \geqslant \delta_o$ and $N_s(IM_o)$ is
the total number of points that fall into the bin $\bra{IM_o -h,IM_o +h}$.

We note that the bMCS approach bears similarities to the stripe analysis introduced by Shome et al. \cite{Shome1998}. However, when using stripe analysis, one scales \emph{all} ground motions to the intensity level of interest. As a result, certain signals are scaled with factors that are considerably larger or smaller than unity, which may lead to gross approximations of the corresponding responses \cite{Luco2007,Cimellaro2009,Mehdizadeh2012}. The reader is referred to \cite{Mehdizadeh2012} for some illustrations of the effects of the scale factor on the introduced bias, with the latter represented by the ratio of the median nonlinear response of the considered system subject to the scaled motions to the respective median response of the system subject to natural motions with all motions characterized by the same IM level. In general, the bias ratio tends to become larger with increasing deviation of the scale factor from unity. On the other hand, the scaling in binned MCS is confined in the vicinity of the intensity level $IM_o$, where the vicinity is defined by the bin width $2h$ chosen so that the scale factors are close to unity. Accordingly, the bias due to ground motion scaling is negligible in bMCS.

Following the above discussion, it should be noted that bias from scaling can be avoided by a proper selection of ground motions. For instance, Shome et al. \cite{Shome1998} showed that the scaling of motions that correspond to a narrow interval of earthquake magnitudes and source-to-site distances does not introduce bias into the nonlinear response estimates. Furthermore, Luco and Bazzurro \cite{Luco2007} showed that the bias can be reduced by selecting records that have appropriate response spectrum shapes. According to Bazzurro et al. \cite{Bazzurro1998} and Vamvatsikos and Cornell \cite{Vamvatsikos2002}, the existence of scale-induced bias also depends on several other factors, such as the structural characteristics and the considered intensity and damage measures. The topic of ground motion scaling is complex and falls outside the scope of this paper. We underline that by using the bMCS approach, we avoid introducing bias in the results independently of the ground motion characteristics or other factors. In the following case studies, the resulting fragility curves serve as reference for assessing the accuracy of the various considered techniques for fragility estimation.

\subsection{Kernel density estimation}
\label{sec3.3}
The fragility function defined in \eqrefe{eq7} may be reformulated using the conditional probability density function (PDF) $f_{\Delta|IM}$ as follows:
\begin{equation}
	\text{Frag}(a;\,\delta_o)= \Prob{\Delta \geq \delta_o | IM=a}=\int\limits_{ \delta_o}^{+\infty} f_{\Delta}(\delta| IM=a) \, \di \delta.
	\label{eq10a}
\end{equation}
By definition, this conditional PDF is given as:
\begin{equation}
	f_{\Delta}(\delta|IM=a)=\frac{f_{\Delta,IM}(\delta,a)}{f_{IM}(a)}
	\label{eq10},
\end{equation}
where $f_{\Delta,IM}(\cdot)$ is the joint distribution of the 
vector $(\Delta,\,IM)$ and $f_{IM}(\cdot)$ is the marginal distribution of the $IM$. If these quantities were known, the fragility function in \eqrefe{eq10a} would be obtained by a mere integration. In this section, we propose to determine the joint and marginal PDFs from a sample set $\left\{ \prt{IM_i,\Delta_i} \right.$, $\left. i=1 \enu N \right\}$ by means of \emph{kernel density estimation} (KDE).

For a single random variable $X$ for which a sample set $\acc{x_1 \enu x_N}$ is available, the kernel density estimate of the PDF reads \cite{WandJones}:
\begin{equation}
	\hat{f}_X\prt{x}=\dfrac{1}{N h} \sum_{i=1}^N K\prt{\dfrac{x-x_i}{h}},
\end{equation}
where $h$ is the \emph{bandwidth} parameter and $K(\cdot)$ is the \emph{kernel} function which integrates to one. Classical kernel functions are the Epanechnikov, uniform, normal and triangular 
functions. The choice of the kernel is known not to affect strongly the quality of the estimate provided the sample set is large enough \cite{WandJones}. In case a standard normal PDF is adopted for the kernel, the kernel density estimate rewrites:
\begin{equation}
	\hat{f}_X\prt{x}=\dfrac{1}{N h} \sum_{i=1}^N \dfrac{1}{\prt{2\pi}^{1/2}} \expo{-\dfrac{1}{2}\prt{\dfrac{x-x_i}{h}}^2}.
	\label{eq12}
\end{equation}
In contrast, the choice of the bandwidth $h$ is crucial since an inappropriate value of $h$ can lead to an oversmoothed or undersmoothed PDF estimate \cite{DuongThesis2004}.

Kernel density estimation may be extended to a random vector $\ve{X} \in \Rr^d$ given an i.i.d sample $\acc{\ve{x}_1 \enu \ve{x}_N}$ \cite{WandJones}:
\begin{equation}
	\hat{f}_{\ve{X}}\prt{\ve{x}}=\dfrac{1}{N  \abs{\mat{H}}^{1/2}} \sum_{i=1}^N K\prt{\mat{H}^{-1/2}(\ve{x}-\ve{x}_i)},
\end{equation}
where $\mat{H}$ is a symmetric positive definite \emph{bandwidth matrix} with determinant denoted by $\abs{\mat{H}}$. When a multivariate standard normal kernel is adopted, the joint distribution estimate becomes:
\begin{equation}
	\hat{f}_{\ve{X}}\prt{\ve{x}}=\dfrac{1}{N \abs{\mat{H}}^{1/2}} \sum_{i=1}^N \dfrac{1}{\prt{2\pi}^{d/2}}\expo{-\frac{1}{2} {(\ve{x}-\ve{x}_i)}\tr \mat{H}^{-1} (\ve{x}-\ve{x}_i)},
	\label{eq14}
\end{equation}
where $(\cdot)\tr$ denotes the transposition. For multivariate problems (\ie $\ve{X} \in \Rr ^d$), the bandwidth matrix typically belongs to one of the following classes: spherical, ellipsoidal and full matrix, which respectively contain 1, $d$ and $d(d+1)/2$ independent unknown parameters. The matrix $\mat{H}$ can be computed by means of plug-in or cross-validation estimators. Both estimators aim at minimizing the asymptotic mean integrated squared error (MISE):
\begin{equation}
	\text{MISE} = \Esp{ \int\limits_{\Rr^d}^{} \bra{ \hat{f}_{\ve{X}}(\ve{x}; \, \mat{H}) - f_{\ve{X}}(\ve{x}) }^2 \di \ve{x} }.
	\label{eqMISE}
\end{equation}
However, the two approaches differ in the formulation of the numerical approximation of MISE. For further details, the reader is referred to Duong \cite{DuongThesis2004}. In the most general case when the correlations between the random variables are not known, the full matrix should be used. In this case, the \textit{smoothed cross-validation estimator} is the most reliable among the cross-validation methods \cite{Duong2005}.

\eqrefe{eq12} is used to estimate the marginal PDF of the $IM$, namely $\hat{f}_{IM}(a)$, from a sample $\acc{IM_i,\,i=1 \enum N}$:
\begin{equation}
	\hat{f}_{IM}(a)=\dfrac{1}{\prt{2\pi}^{1/2} N h_{IM}} \sum_{i=1}^N \expo{-\dfrac{1}{2}\prt{\dfrac{a-{IM}_i}{h_{IM}}}^2}.
\end{equation}
\eqrefe{eq14} is used to estimate the joint PDF $\hat{f}_{\Delta,IM}(\delta,a)$ from the data pairs $\acc{(IM_i,\,\Delta_i),\,i=1 \enum N}$:
\begin{equation}
	\hat{f}_{\Delta,IM}(\delta,a)= \dfrac{1}{2\pi N \abs{\mat{H}}^{1/2}}   \sum\limits_{i=1}^N \expo{-\dfrac{1}{2} {\begin{pmatrix} \delta-\Delta_i\\ a-IM_i \end{pmatrix}}\tr \mat{H}^{-1}  \begin{pmatrix} \delta-\Delta_i\\ a-IM_i \end{pmatrix} }.
	\label{eq15}
\end{equation}
The conditional PDF $f_{\Delta}(\delta|IM=a)$ is eventually estimated by plugging the estimations of the numerator and denominator in \eqrefe{eq10}. The proposed estimator of the fragility function eventually reads:
\begin{equation}
	\widehat{\text{Frag}}(a;\,\delta_o)
	=  \dfrac{h_{IM}}{\prt{2\pi \abs{\mat{H}}}^{1/2}} \dfrac  {  \int\limits_{ \delta_o}^{+\infty}  \sum\limits_{i=1}^N \expo{-\dfrac{1}{2} {\begin{pmatrix} \delta-\Delta_i\\ a-IM_i \end{pmatrix}}\tr \mat{H}^{-1}  \begin{pmatrix} \delta-\Delta_i\\ a-IM_i \end{pmatrix} } \text{d}\delta}
	{ \sum\limits_{i=1}^N \expo{-\dfrac{1}{2}\prt{\dfrac{a-{IM}_i}{h_{IM}}}^2}}.
	\label{eq17}
\end{equation}

The choice of the bandwidth parameter $h$ and the bandwidth matrix $\mat{H}$ plays a crucial role in the estimation of fragility curves, as seen in \eqrefe{eq17}. In the above formulation, the same bandwidth is considered for the whole range of the $IM$ values. However, there are typically few observations available corresponding to the upper tail of the distribution of the $IM$. This is due to the fact that the annual frequency of seismic motions with $IM$ values in the respective range (\eg $PGA$ exceeding $1g$) is low (see \eg \cite{Frankel2000}). This is also the case when synthetic ground motions are used, since these are generated consistently with statistical features of recorded motions. Preliminary investigations have shown that by applying the KDE method on the data in the original scale, the fragility curves for the higher demand thresholds tend to be unstable in their upper tails \cite{SudretMaiCFM2013}. To reduce effects from the scarcity of observations at large $IM$ values, we propose the use of KDE in the logarithmic scale, as described next.

Let us consider two random variables $X$, $Y$ with positive supports, and their logarithmic transformations  $U= \ln X$ and $V= \ln Y$.
One has:
\begin{equation}
	\int\limits_{y_0}^{+\infty} f_Y(y | X=x) \, \di y 
	=
	\int\limits_{y_0}^{+\infty} \dfrac{f_{X,Y}(x,y)}{f_X(x)} \, \di y 
	= 
	\int\limits_{\ln y_0}^{+\infty} \dfrac{\dfrac{f_{U,V}(u,v)}{x\,y}}{\dfrac{f_U(u)}{x}} \, y \, \di v
	=
	\int\limits_{\ln y_0}^{+\infty} f_V (v | U=u) \, \di v.
	\label{eq26}
\end{equation}
Accordingly, by substituting $X =IM$ and $Y=\Delta$, the fragility function in \eqrefe{eq10a} can be obtained in terms of $U=\ln IM$ and $V=\ln \Delta$ as:
\begin{equation}
	\widehat{\text{Frag}}(a;\,\delta_o)=\int\limits_{ \delta_o}^{+\infty} \hat{f}_{\Delta}(\delta| IM=a) \, \di \delta
	= \int\limits_{\ln \delta_o}^{+\infty} \hat{f}_V (v | U= \ln a) \, \di v.
	\label{eq27}
\end{equation}

The use of a constant bandwidth in the logarithmic scale is equivalent to the use of a varying bandwidth in the original scale, with larger bandwidths corresponding to larger values of $IM$. The resulting fragility curves are smoother than those obtained by applying KDE with the data in the original scale.

\subsection{Epistemic uncertainty of fragility curves}
\label{sec3.4}

It is of major importance in fragility analysis to investigate the variability in the estimated curves arising due to epistemic uncertainty. This is because a fragility curve is always computed based on a limited amount of data, \ie a limited number of ground motions and related structural analyses. Large epistemic uncertainties may affect significantly the total variability of the seismic risk assessment outcomes. Characterizing and propagating epistemic uncertainties in seismic loss estimation has therefore attracted attention from several researchers \cite{Baker2008a,Bradley2010,Liel2009}.

The theoretical approach to determine the variability of an estimator relies on repeating the estimation with an ensemble of different random samples. However, this approach is not feasible in earthquake engineering because of the high computational cost. In this context, the \textit{bootstrap resampling} technique is deemed appropriate \cite{Baker2008a}. Given a set of observations $\cx=\prt{\ve{X}_1 \enum \ve{X}_n}$ of $\ve X$ following an unknown probability distribution, the bootstrap method allows estimation of the statistics of a random variable that depends on $\ve X$ in terms of the observed data $\cx$ and their empirical distribution \cite{Efron1979}.

To estimate statistics of the fragility curves with the bootstrap method, we first draw $M$ independent random samples \emph{with replacement} from the original data set $\acc{\prt{IM_i,\Delta_i}, i=1 \enu N}$. These represent the so-called bootstrap samples. Each bootstrap sample has the same size $N$ as the original sample, but the observations are different: in a particular sample, some of the original observations may appear multiple times while others may be missing. Next, we compute the fragility curves for each bootstrap sample using the approaches in Sections \ref{sec3.1}, \ref{sec3.2} and \ref{sec3.3}. Finally, we perform statistical analysis of the so-obtained $M$ bootstrap curves. In the subsequent example illustration, the above procedure is employed to evaluate the median and 95\% confidence intervals of the estimated fragility curves and also, to assess the variability of the $IM$ value corresponding to a $50 \%$ probability of failure.

\section{Synthetic ground motions}
\label{sec2}
\vspace{-2pt}
\subsection{Properties of recorded ground motions}
\label{sec2.1}
Let us consider a recorded earthquake accelerogram $a(t)$, $t\in [0,T]$ where $T$ is the total duration of the motion. The peak ground acceleration is $PGA= \max\limits_{\substack{t\in[0,T]}} \, \lvert a(t) \lvert $. The Arias intensity $I_a$ is defined as:
\begin{equation}
	I_a=\frac{\pi}{2g} \int\limits_0^T {a^2(t)}\, \mathrm{d} t.
	\label{eq:Ia}
\end{equation}
Defining the cumulative square acceleration as:
\begin{equation}
	I(t)=\frac{\pi}{2g} \int\limits_0^t \,{a^2(\tau)}\, \mathrm{d} \tau \, ,
	\label{eq:It}
\end{equation}
one determines the time instant $t_{\alpha}$ by:
\begin{equation}
	t_{\alpha}: \qquad I(t_{\alpha})=\alpha I_a \qquad \alpha \in [0,1].
\end{equation}
In addition to the Arias intensity, important properties of the accelerogram in the time domain include the effective duration, defined as $D_{5-95}=t_{95\%} - t_{5\%}$, and the instant $t_{mid}$ at the middle of the strong-shaking phase \cite{Rezaeian2010}. Based on the investigation of a set of recorded ground motions, Rezaeian and Der~Kiureghian~\cite{Rezaeian2010} proposed that  $t_{mid}$ is taken as the time when $45\%$ of the Arias intensity is reached \ie  $t_{mid}\equiv t_{45\%}$.

Other important properties of the accelerogram are related to its frequency content. Analyses of recorded ground motions indicate that the time evolution of the predominant frequency of an accelerogram can be represented by a linear model, whereas its bandwidth can be considered constant~\cite{Rezaeian2008}. Rezaeian and Der~Kiureghian~\cite{Rezaeian2008} describe the evolution of the predominant frequency in terms of its value $\omega_{mid}$ at the time instant $t_{mid}$ and the slope of the evolution  $\omega'$. The same authors describe the bandwidth in terns of the bandwidth parameter $\zeta$. A procedure for estimating the parameters $\omega_{mid}$, $\omega'$ and $\zeta$ for a given accelerogram is presented in \cite{Rezaeian2008}, whereas a simplified version is proposed in \cite{Rezaeian2010}.

Next, we describe a method for simulating synthetic accelerograms in terms of the set of parameters $\left(I_a, D_{5-95}, t_{mid}, \omega_{mid}, \omega', \zeta_f\right)$; this method will be used to generate the seismic motions in a subsequent case study.

\subsection{Simulation of synthetic ground motions}

The use of synthetic ground motions has been attracting an increasing interest from the earthquake engineering community. This practice overcomes the limitations posed by the small number of records typically available for a design scenario and avoids the need to scale the motions. Use of synthetic ground motions allows one to investigate the structural response for a large number of motions, which is nowadays feasible with the available computer resources (see \eg \cite{Kwong2015}).

Different stochastic ground motion models can be found in the literature, which can be classified in three types \cite{Rezaeian2010}: record-based parameterized models that are fit to recorded motions, source-based models that consider the physics of the source mechanism and wave travel-path, and hybrid models that combine elements from both source- and record-based models. Vetter and Taflanidis \cite{Vetter2014} compared the source-based model by Boore \cite{Boore2003} with the record-based model by Rezaeian and Der Kiureghian \cite{Rezaeian2010} with respect to the estimated seismic risks. It was found that the latter leads to higher estimated risks for low-magnitude events, but the risks are quantified in a consistent manner exhibiting correlation with the hazard characteristics. This model is employed in the present study to generate a large suite of synthetic ground motions that are used to obtain pairs of the ground motion intensity measure and the associated structural response, $(IM, \Delta)$, in order to conduct fragility analysis. The approach, originally proposed in \cite{Rezaeian2008}, is summarized below.

The seismic acceleration $a(t)$ is represented as a non-stationary
process. In particular, the non-stationarity is separated into
two components, namely a spectral and a temporal one, by means of a
modulated filtered Gaussian white noise:
\begin{equation} 
	a(t)=  \frac{q(t,\ve{\alpha}) }{\sigma_h{(t)}}
	\int\limits_{0}^t \,{h \bra{ t-\tau,\ve{\lambda}\prt{\tau} }
		\omega(\tau)}\, \mathrm{d} \tau,
	\label{eq:at}
\end{equation} in which $q(t,\ve{\alpha})$ is the deterministic
non-negative \emph{modulating function}, the integral is the non-stationary response of a linear filter subject to a Gaussian white-noise excitation and $\sigma_h{(t)}$ is the standard deviation
of the response process. The Gaussian white-noise process denoted by $\omega(\tau)$ will pass through a filter $h \bra{t-\tau,\ve{\lambda}(\tau)}$, which is selected as the pseudo-acceleration response of a single-degree-of-freedom (SDOF) linear oscillator:
\begin{equation}
	\begin{aligned}
		& h \bra{t-\tau,\ve{\lambda}(\tau)}=0 \quad \text{for} \quad t < \tau \\ 
		& h \bra{t-\tau,\ve{\lambda}(\tau)}=
		\frac{\omega_f(\tau)}{\sqrt{1-\zeta_f^2(\tau)}} \mathrm{exp}
		\bra{-\zeta_f(\tau) \omega_f(\tau) (t-\tau) }  \sin \bra{
			\omega_f(\tau) \sqrt{1-\zeta_f^2(\tau)} (t-\tau) } \\
		& \quad \quad \quad \quad \quad \quad \quad \quad \text{for}
		\quad t \geq \tau.
	\end{aligned}
\end{equation} 
In the above equation, $\ve{\lambda}(\tau)=\prt{ \omega_f(\tau),\zeta_f(\tau)}$ is the vector of time-varying parameters of the filter $h$, with $\omega_f(\tau)$ and $\zeta_f(\tau)$ respectively denoting the filter's natural frequency and damping ratio at instant $\tau$. Note that $\omega_f(\tau)$ corresponds to the evolving predominant frequency of the ground motion represented, while $\zeta_f(\tau)$ corresponds to the bandwidth parameter of the motion. As noted in Section~\ref{sec2.1}, $\zeta_f(\tau)$ may be taken as a constant $\prt{ \zeta_f(\tau) \equiv \zeta }$, while $\omega_f(\tau)$ is approximated as a linear function:
\begin{equation} 
	\omega_f(\tau) =\omega_{mid} + \omega'(\tau- t_{mid}),
	\label{eq3}
\end{equation} 
where $ t_{mid}$, $\omega_{mid}$ and $\omega'$ are as defined in Section~\ref{sec2.1}. After being normalized by the standard deviation $\sigma_h{(t)}$, the integral in \eqrefe{eq:at} becomes a unit-variance process with time-varying frequency and constant bandwidth. The non-stationarity in intensity is then captured by the modulating function $q(t,\ve{\alpha})$, which determines the shape, intensity and duration $T$ of the signal. This is typically described by a Gamma-like function \cite{Rezaeian2010}:
\begin{equation}
	q(t,\ve{\alpha})=\alpha_1 t^{\alpha_2-1}\mathrm{exp}(-\alpha_3t),
\end{equation}
where $\ve{\alpha}=\acc{\alpha_1,\alpha_2,\alpha_3}$ is
directly related to the energy content of the signal through the quantities $I_a$, $D_{5-95}$ and $t_{mid}$ defined in Section \ref{sec2.1} (see \cite{Rezaeian2010} for details).

For computational purposes, the acceleration in \eqrefe{eq:at} can be discretized as follows:
\begin{equation} 
	\hat{a}(t)=q(t,\ve{\alpha})\sum_{i=1}^n s_i \prt{
		t,\ve{\lambda}(t_i)} \, U_i,
	\label{eq:at2}
\end{equation} where the standard normal random variable $U_i$
represents an impulse at instant $t_i=i \times \dfrac{T}{n} \, , \, i=1
\enu n $, ($T$ is the total duration) and $s_i(t,\ve{\lambda}(t_i))$ is
given by:
\begin{equation}
	s_i(t,\ve{\lambda}(t_i)) = \dfrac{h\bra{t-t_i,\ve{\lambda}(t_i)} }{\sqrt{\sum_{j=1}^i h^2 \bra{t-t_j,\ve{\lambda}(t_j)} }}.
\end{equation}
As a summary, the considered seismic motion generation model consists of the three temporal parameters $\left(\alpha_1,\alpha_2,\alpha_3\right)$, which are related to $\left(I_a, D_{5-95}, t_{mid} \right)$, the three spectral parameters $\left(\omega_{mid}, \omega', \zeta_f\right) $ and the standard Gaussian random vector $\ve{U}$ of size $n$. Rezaeian and Der Kiureghian \cite{Rezaeian2010} proposed a methodology for determining the temporal and spectral parameters according to earthquake and site characteristics, \ie the type of faulting of the earthquake (strike-slip fault or reverse fault), the closest distance from the recording site to the ruptured area and the shear-wave velocity of the top 30~m of the site soil. For the sake of simplicity, in this paper these parameters are directly generated from the statistical models given in \cite{Rezaeian2010}, which are obtained from analysis of a large set of recorded ground motions.

\section{Steel frame structure subject to synthetic ground motions}
\label{sec4}
\subsection{Problem setup}
\begin{figure}[!ht]
	\centering
	\subfigure
	{
		\includegraphics[width=0.49\textwidth]{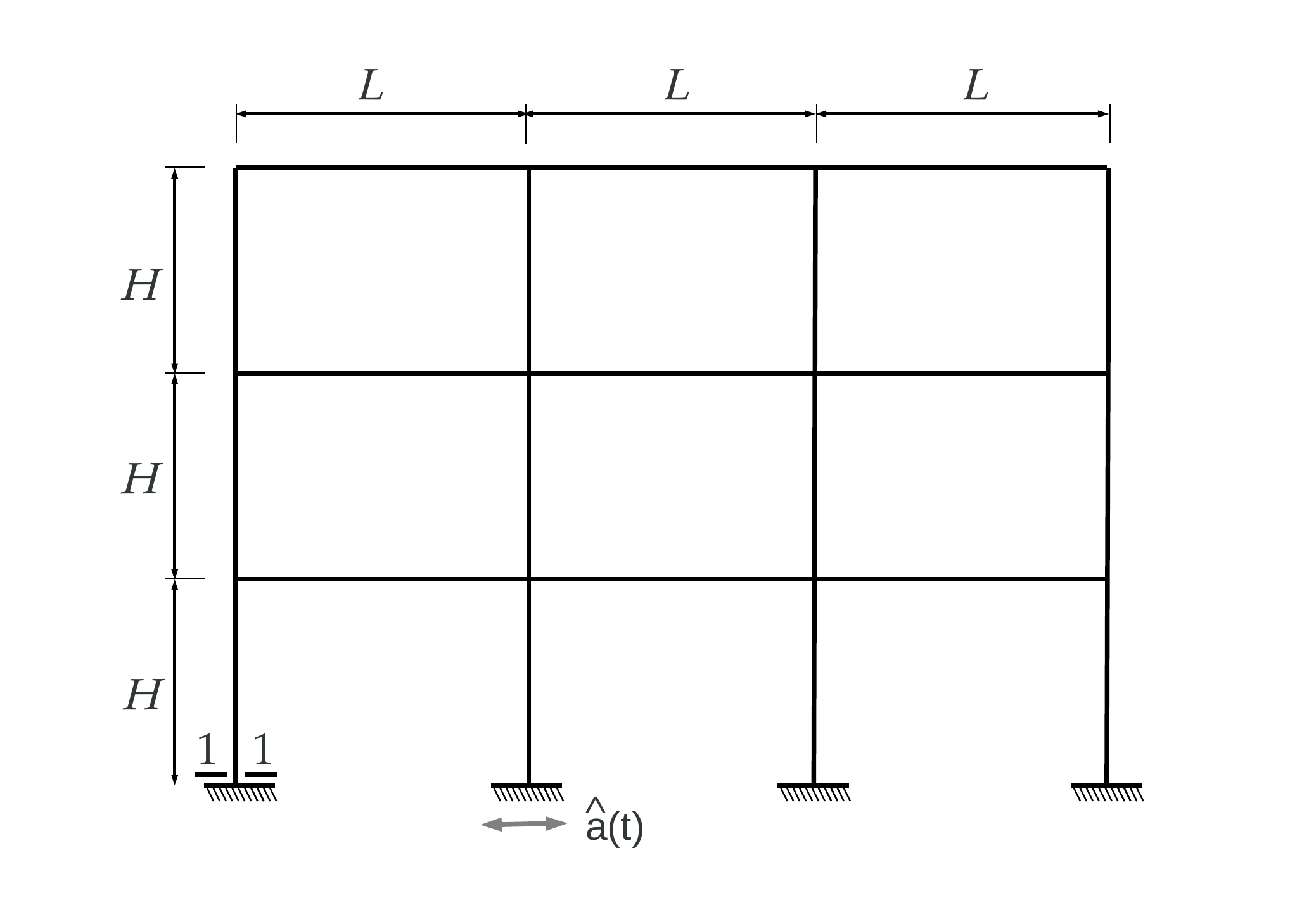}
	}
	\subfigure
	{
		\includegraphics[width=0.45\textwidth]{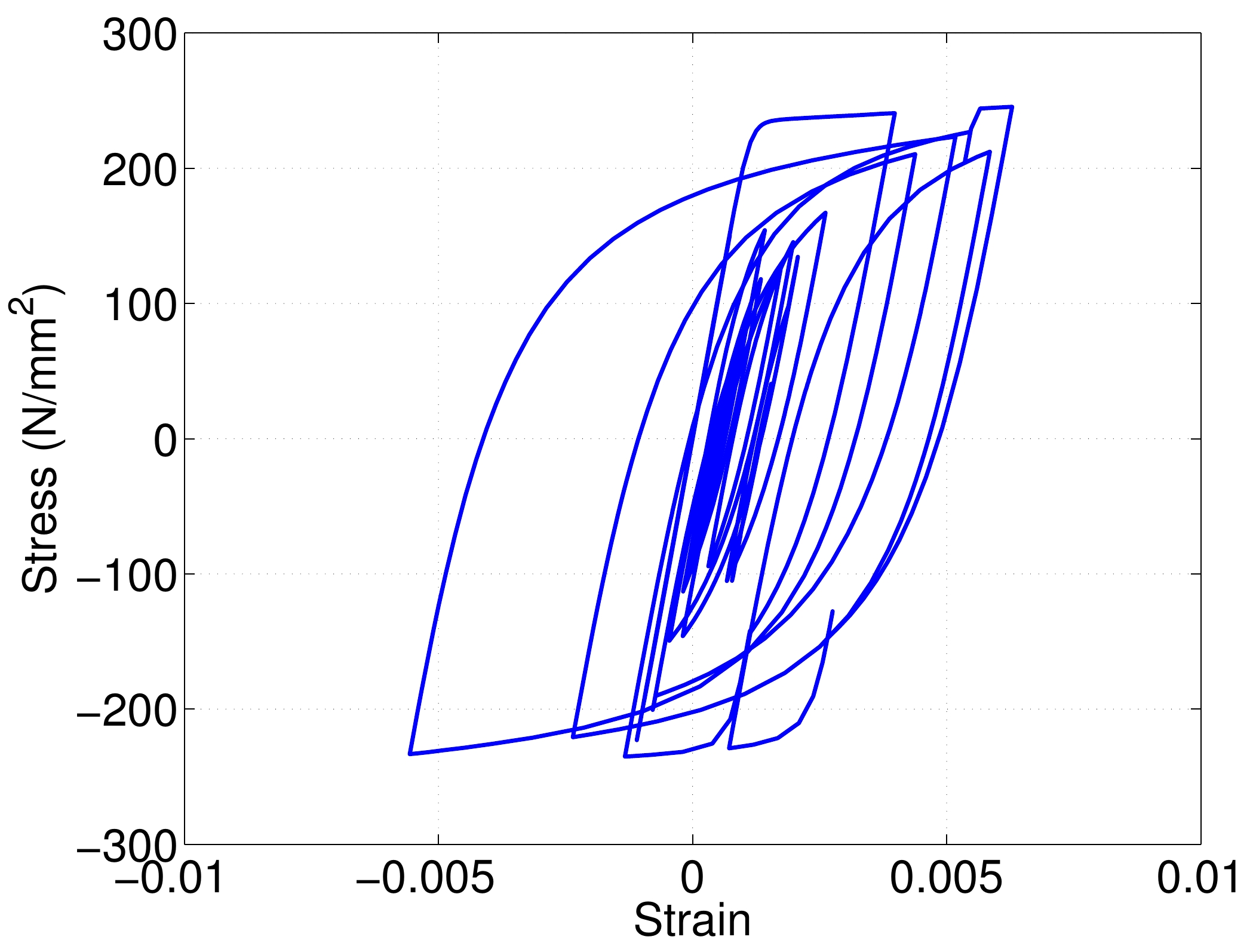}
	}
	\caption{(Left) Steel frame structure. (Right) Hysteretic behavior of steel material at section 1-1 for an example ground motion.}
	\label{structure}
\end{figure}

We determine the fragility curves for the three-storey three-span steel frame shown in \figref{structure}. The dimensions of the structure are: storey-height $H=3$~m, span-length $L=5$~m. The vertical load consists of dead load (weight of frame elements and supported floors) and live load (in accordance with Eurocode~1~\cite{EC1}) resulting in a
total distributed load on the beams $q=20$~kN/m. In the preliminary design stage, the standard European I beams with designation IPE 300 A and IPE 330 O are chosen respectively for the beams and columns. The steel material has a nonlinear isotropic hardening behavior following the uniaxial Giuffre-Menegotto-Pinto steel model as implemented in the finite element software OpenSees \cite{OpenSees}. Ellingwood and Kinali \cite{Ellingwood2009} have shown that uncertainty in the properties of the steel material has a negligible effect on seismic fragility curves. Therefore, the mean material properties are used in the subsequent fragility analysis: $E_0= 210,000$~MPa for the Young's modulus (initial elastic tangent in the stress-strain curve), $f_y=264$~MPa for the yield strength \cite{EC3,jcss} and $b=0.01$ for the strain hardening ratio (ratio of post-yield to initial tangent in the stress-strain curve). \figref{structure} depicts the hysteretic behavior of the steel material at a specified section for an example ground motion. The structural components are modelled with nonlinear force-based beam-column elements characterized by distributed plasticity along their lengths, while use of fiber sections allows modelling the plasticity over the element cross-sections \cite{Deierlein2010}. The connections between structural elements are modeled with rigid nodes. The first two natural periods of the building obtained by modal analysis are $T_1=0.61$~s and $T_2=0.181$~s, corresponding to natural frequencies $f_1=1.64$~Hz and $f_2=5.53$~Hz. Rayleigh damping is considered with the damping ratio of the first two modes set equal to 2\%.

\begin{figure}[!ht]
	\centering
	\subfigure
	{
		\includegraphics[width=0.47\textwidth]{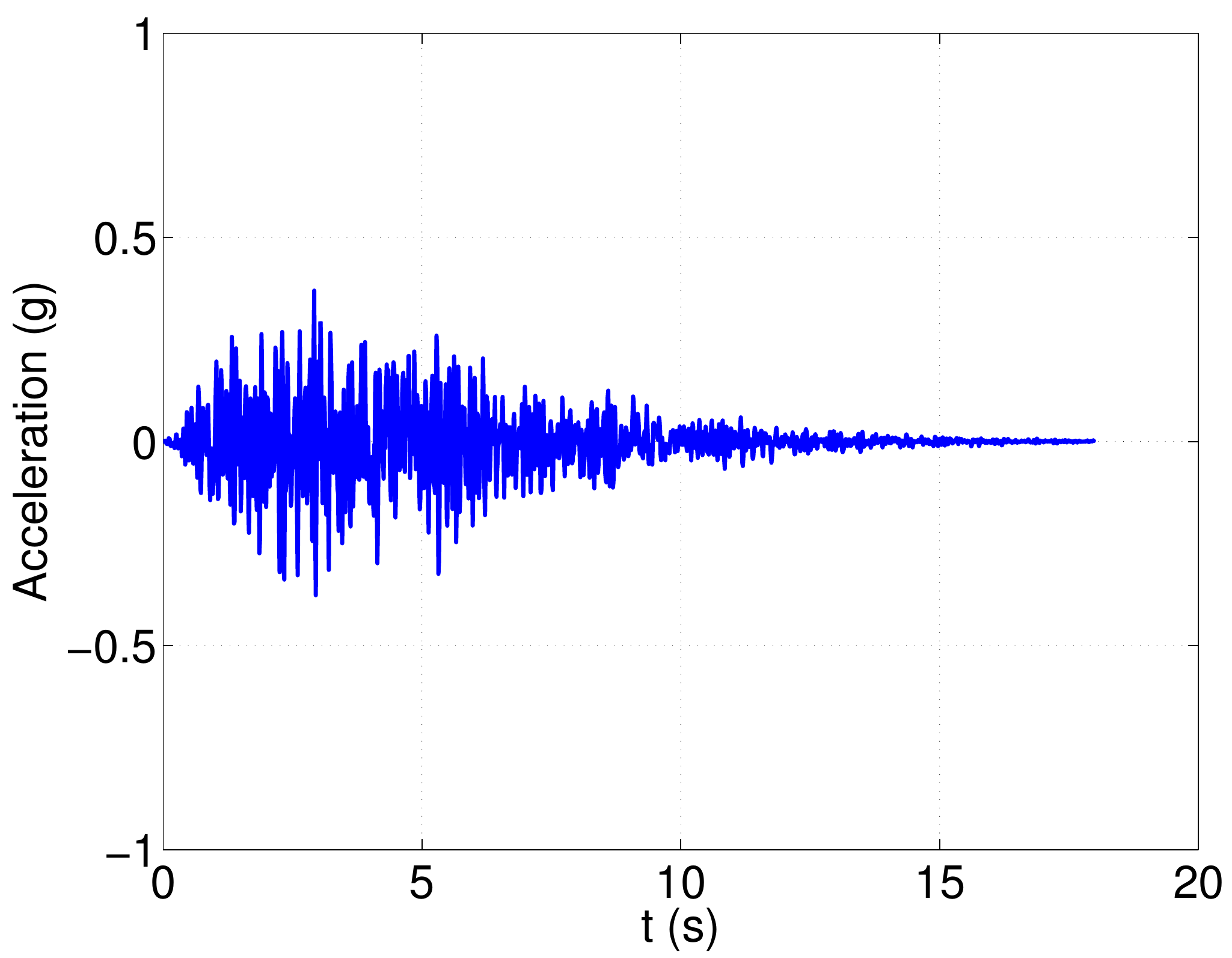}
	}
	\subfigure
	{
		\includegraphics[width=0.47\textwidth]{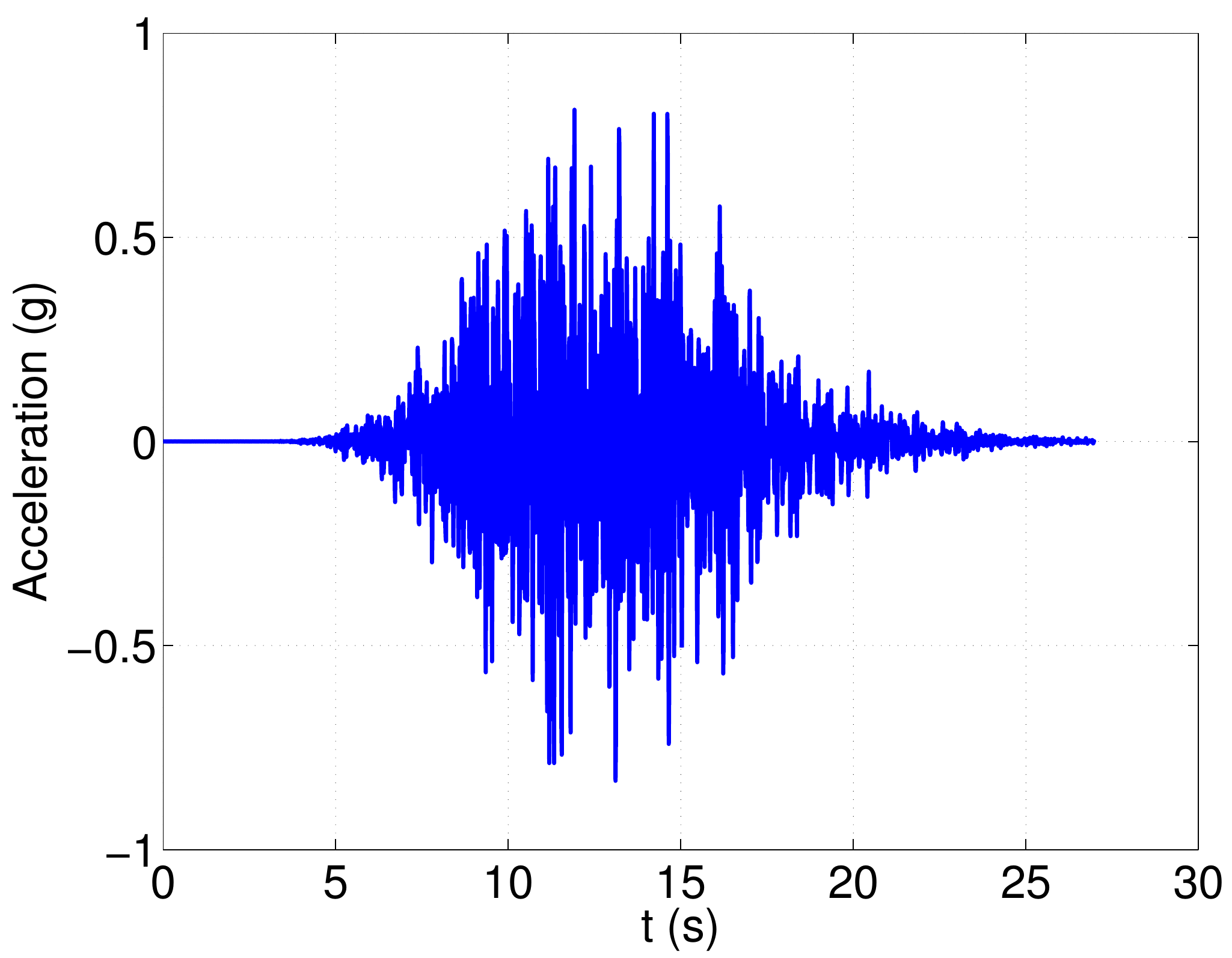}
	}
	\caption{Examples of synthetic ground motions.}
	\label{fig4.1.3}
\end{figure}

The structure is subject to seismic motions represented by synthetic acceleration time histories at the ground level. Each time history is modelled in terms of six randomized parameters $\prt{\alpha_1,\alpha_2,\alpha_3,\omega_{mid}, \omega', \zeta_f }$ 
directly related to the parameters in Table~\ref{tab:1} and a Gaussian input vector $\ve{U}$ as described in Section~\ref{sec2}.  The statistics of the parameters in Table~\ref{tab:1} are taken from \cite{Rezaeian2010};  in the latter study, the authors derived the listed distributions and associated parameters by analyzing a set of recorded ground motions corresponding to strong strike-slip and reserve earthquakes with moment magnitudes in the range 6-8 and rupture distances in the range of 10-100~km. The reader is referred to \cite{Rezaeian2010} for viewing the correlations between these parameters. The duration of each time history is computed from the corresponding set of parameters $\prt{\alpha_1,\alpha_2,\alpha_3}$ and is used to determine the size of the Gaussian vector $\ve{U}$. Two example synthetic acceleration time histories are shown in \figref{fig4.1.3}. Transient dynamic analyses of the frame are carried out for a total of $N=20,000$ synthetic motions using the finite element software OpenSees.

\begin{table}[!ht]
	\caption{Statistics of synthetic ground motion parameters according to~\cite{Rezaeian2010}. }
	\centering
	\begin{tabular}{m{0.2\textwidth} m{0.3\textwidth} m{0.12\textwidth} m{0.12\textwidth} m{0.12\textwidth}}
		\toprule
		Parameter & Distribution & Support & $\mu_X$ & $\sigma_X$ \\
		\midrule
		$I_a$ (s$\times$g) & Lognormal & (0, +$\infty$) & 0.0468 & 0.164 \\
		$D_{5-95}$ (s) & Beta      & [5, 45]  & 17.3 & 9.31 \\
		$t_{mid}$ (s) & Beta & [0.5, 40] & 12.4 & 7.44 \\
		$\omega_{mid}$/{2$\pi$} (Hz) & Gamma & (0, +$\infty$) & 5.87 & 3.11 \\
		$\omega'$/{2$\pi$} (Hz) & Two-sided exponential&[-2, 0.5] &-0.089 & 0.185 \\
		$\zeta_f$ &Beta & [0.02, 1] & 0.213 & 0.143\\
		\bottomrule
	\end{tabular}
	\label{tab:1}
\end{table}

Numerous types of $IM$ can be used to describe the earthquake severity, see \eg \cite{Mackie2004}. Peak ground acceleration ($PGA$) is a convenient measure that is straightforward to obtain from a given time history and has been traditionally used in attenuation relationships and design codes. However, structural responses may exhibit large dispersions for a certain $PGA$, since they are also highly dependent on other features of earthquake motions, e.g. the frequency content and duration of the strong motion phase. Structure-specific $IM$s, such as the spectral acceleration $Sa$ and the pseudo spectral acceleration $Psa$, tend to be better correlated with structural responses \cite{Mackie2004,Padgett2008a}. In the following, we compute fragility curves considering both $PGA$ and $Sa$ as $IM$s. $Sa$ represents $Sa(T_1)$ \ie the spectral acceleration for a single-degree-of-freedom system with period equal to the fundamental period $T_1$ of the frame and viscous damping ratio equal to 2\%.

The engineering demand parameter commonly considered in fragility analysis of steel buildings is the maximal inter-storey drift ratio, \ie the maximal difference of horizontal displacements between consecutive storeys normalized by the storey height (see \eg \cite{Ellingwood2009,Cornell2002,Lagaros2007}). Accordingly, we herein develop fragility curves for three different thresholds of the maximal inter-storey drift ratio over the frame. To gain insight into structural performance, we consider the thresholds 0.7\%, 1.5\% and 2.5\%, which are associated with different damage states in seismic codes. In particular, the thresholds 0.7\% and 2.5\% are recommended in \cite{FEMA2000} to respectively characterize light and moderate damage for steel frames, while the threshold 1.5\% corresponds to the damage limitation requirement for buildings with ductile non-structural elements according to Eurocode 8 \cite{EC8eng}. These descriptions only serve as rough damage indicators, since the relationship between drift limit and damage in the PBEE framework is probabilistic.

\subsection{Fragility curves}
\begin{figure}[!ht]
	\centering
	\subfigure
	{
		\includegraphics[width=0.45\textwidth]{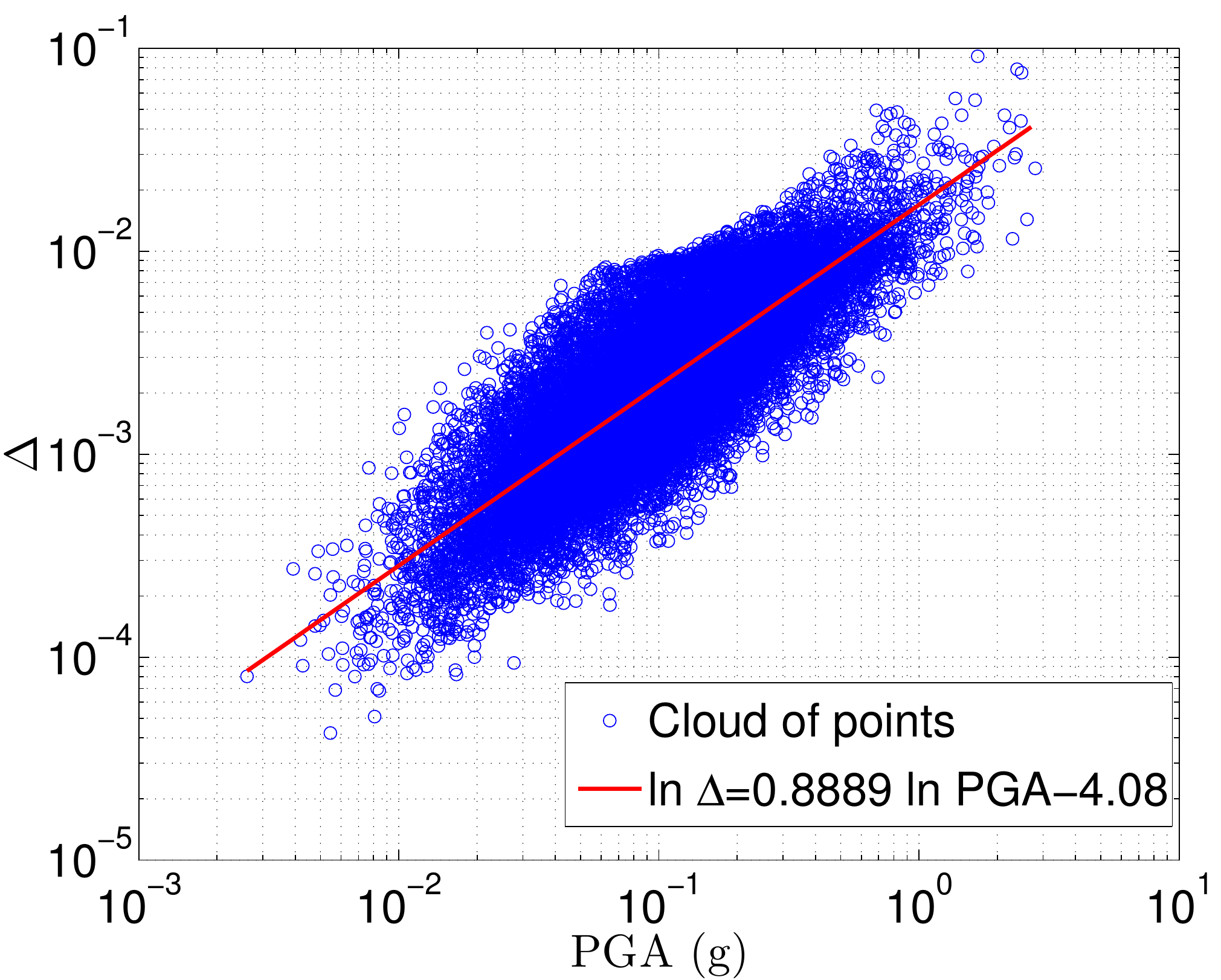}
	}
	\subfigure
	{
		\includegraphics[width=0.45\textwidth]{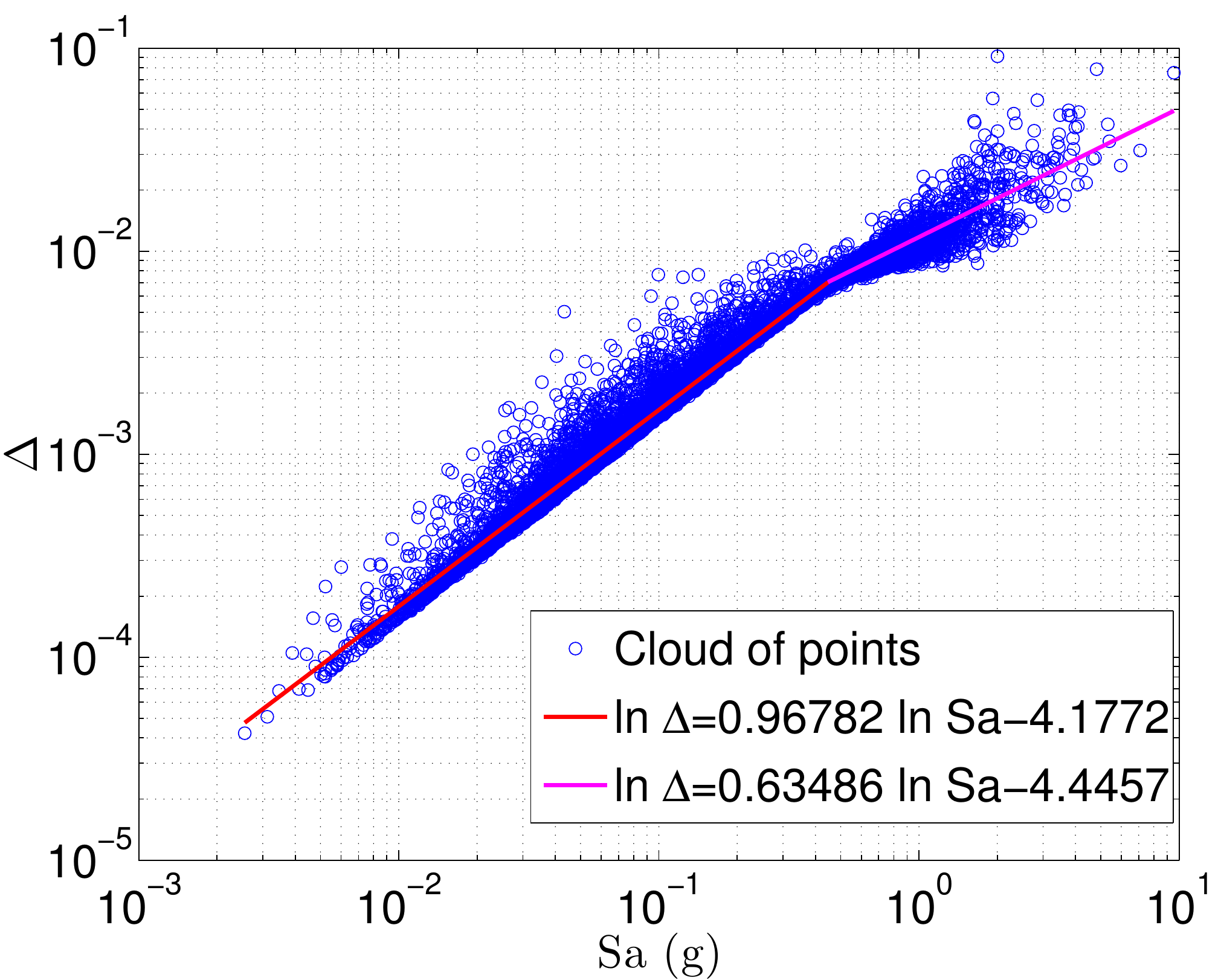}
	}
	\caption{Paired data $\acc{\prt{IM_i,\Delta_i}, i=1 \enu N}$ and fitted models in log-scale (the units of the variables in the fitted models are the same as in the axes of the graphs).}
	\label{fig4.2.1}
\end{figure}

As described in Section \ref{sec3}, the lognormal approach relies on assuming that the fragility curves have the shape of a lognormal CDF and estimating the parameters of this CDF. Using the maximum likelihood estimation (MLE) approach, the observed failures for each drift threshold are modeled as outcomes of a Bernoulli experiment and the parameters ($\alpha$, $\beta$) of the fragility curves are determined by maximizing the respective likelihood function. Using the linear regression (LR) technique, the parameters of the lognormal curves are derived by fitting a linear model to the paired data $(\ln IM,\,\ln \Delta)$. \figref{fig4.2.1} depicts the paired data $(\ln PGA,\,\ln \Delta)$ and $(\ln Sa,\,\ln \Delta)$ together with the fitted models based on linear regression. It can be seen that a single linear model is not appropriate for the cloud of points $(\ln Sa,\,\ln \Delta)$ and thus, bilinear regression is used in this case (see also \cite{Mackie2003,Ramamoorthy2006,Bai2011} for use of a similar model). The break point in the bilinear model ($Sa = 0.45\,g$) is determined according to the method presented in \cite{Muggeo2003} using the \texttt{R} package \texttt{segmented}. When $PGA$ is used as $IM$, the coefficient of determination of the fitted linear model is $R^2=0.663$; when $Sa$ is used as $IM$, it is $R_1^2=0.978$ and $R_2^2=0.785$ for the first and second part of the bilinear model, respectively. Note that use of $Sa$ as $IM$ leads to a smaller dispersion, \ie a smaller $\zeta$ in \eqrefe{eq8}, as compared to $PGA$; this is expected since $Sa$ is a structure-specific $IM$. In the bMCS method, the bin width $h$ is set equal to $0.25\,IM_o$. The resulting scale factors vary in the range $[0.75,\,1.25]$ corresponding to a bias ratio approximately equal to unity \cite{Mehdizadeh2012}. The KDE approach requires estimation of the bandwidth parameter and the bandwidth matrix. Using the cross-validation estimation implemented in \texttt{R} \cite{Duongks2007}, these are determined as $h= 0.133$, $\ve{H}= \begin{bmatrix}
0.031 & 0.024;~
0.024 & 0.027
\end{bmatrix}$
when $PGA$ is used as $IM$, and $h=0.155$, 
$\ve{H}=\begin{bmatrix}
0.023 & 0.023;~
0.023 & 0.024
\end{bmatrix}$ when $Sa$ is used as $IM$. 

For the two types of $IM$ and the three drift limits considered, \tabref{tab:2} lists the medians and log-standard deviations of the lognormal curves obtained with both the MLE and LR approaches. The median determines the position where the curve attains the value 0.5, whereas the log-standard deviation is a measure of the steepness of the curve. Note that the MLE approach yields a distinct log-standard deviation for each drift threshold, whereas a single log-standard deviation for all drift thresholds is obtained with the LR approach. The medians of the KDE-based curves are also computed and are listed in \tabref{tab:2} for comparison. The KDE-based medians, which serve as the reference values, may be overestimated or underestimated by the lognormal approach, depending on the method used to estimate the parameters, the considered $IM$ and the drift threshold; the absolute deviations tend to be larger for larger drift thresholds.

\begin{table}[!ht]
	\caption{Steel frame structure - Parameters of the obtained fragility curves. }
	\centering
	\begin{tabular}{cccccc}
		\toprule
		& & \multicolumn{2}{c}{PGA}  & \multicolumn{2}{c}{Sa}    \\
		\cline{3-6}
		$\delta_o$ & Approach & Median & Log-std & Median & Log-std  \\
		\midrule  
		\multirow{3}{*}{0.7\%}  
		& MLE & $0.35\,g$ &  0.70 & $0.49\,g$ &  0.36  \\
		& LR & $0.37\,g$ & 0.64 & $0.44\,g$ & 0.13 \\
		& KDE & $0.36\,g$ &  & $0.45\,g$ &  \\
		\cline{2-6}
		\multirow{3}{*}{1.5\%}  
		& MLE & $1.10\, g$ &  0.56 & $1.66\,g$ &  0.31   \\
		& LR & $0.87\,g$ & 0.64 & $1.47\,g$ & 0.24  \\
		& KDE & $1.08\,g$ & & $1.53\,g$ &     \\
		\cline{2-6}
		\multirow{3}{*}{2.5\%}  
		& MLE & $1.76\,g$ &  0.56 & $2.82\,g$ &  0.37  \\
		& LR & $1.55\,g$ & 0.64 & $3.29\,g$ & 0.24 \\
		& KDE & $1.82\,g$ & & $3.04\,g$ &     \\
		
		\bottomrule
	\end{tabular}
	\label{tab:2}
\end{table}

\begin{figure}[!ht]
	\centering
	\subfigure
	{
		\includegraphics[width=0.47\textwidth]{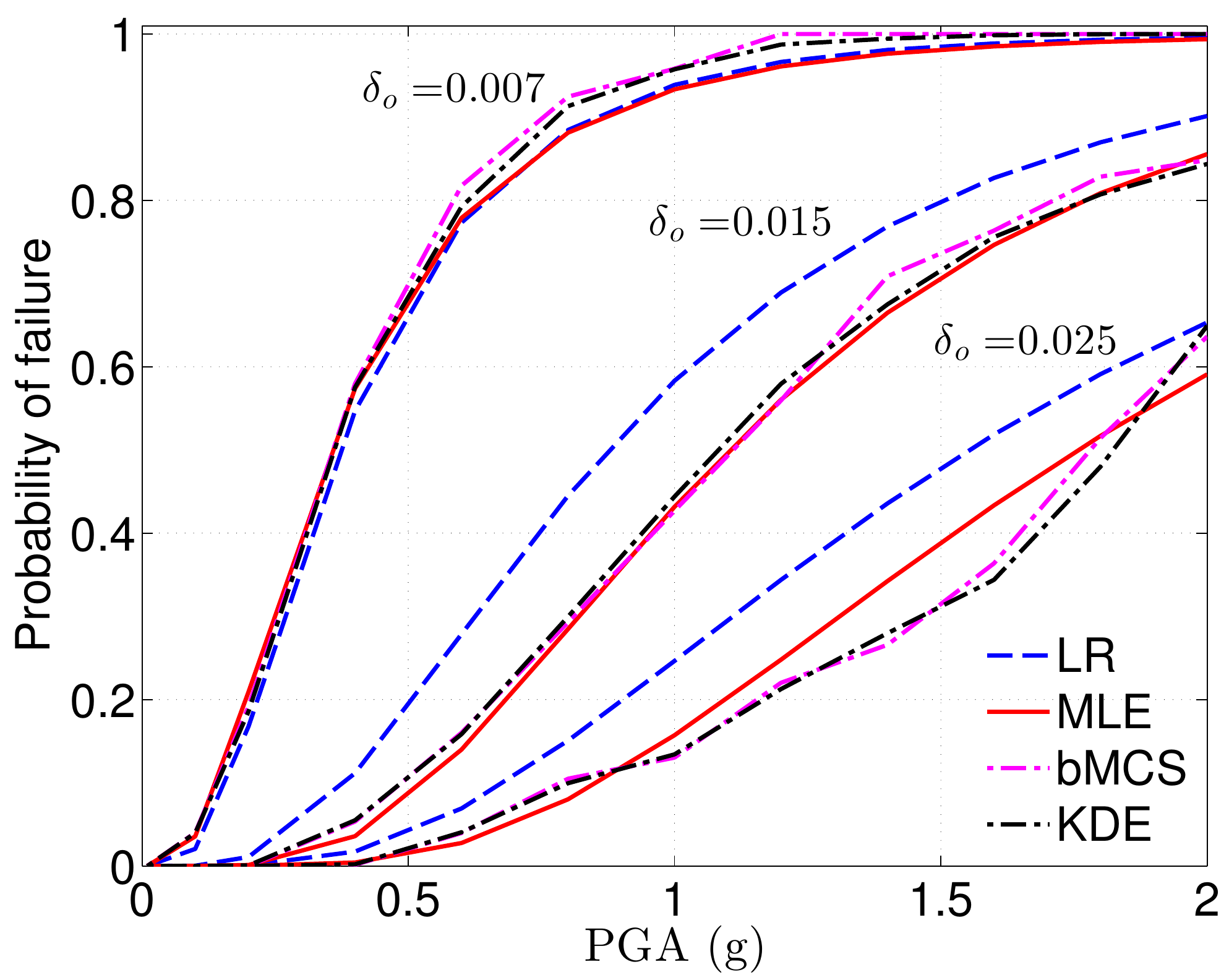}
		\label{fig4.2.2a}
	}
	\subfigure
	{
		\includegraphics[width=0.47\textwidth]{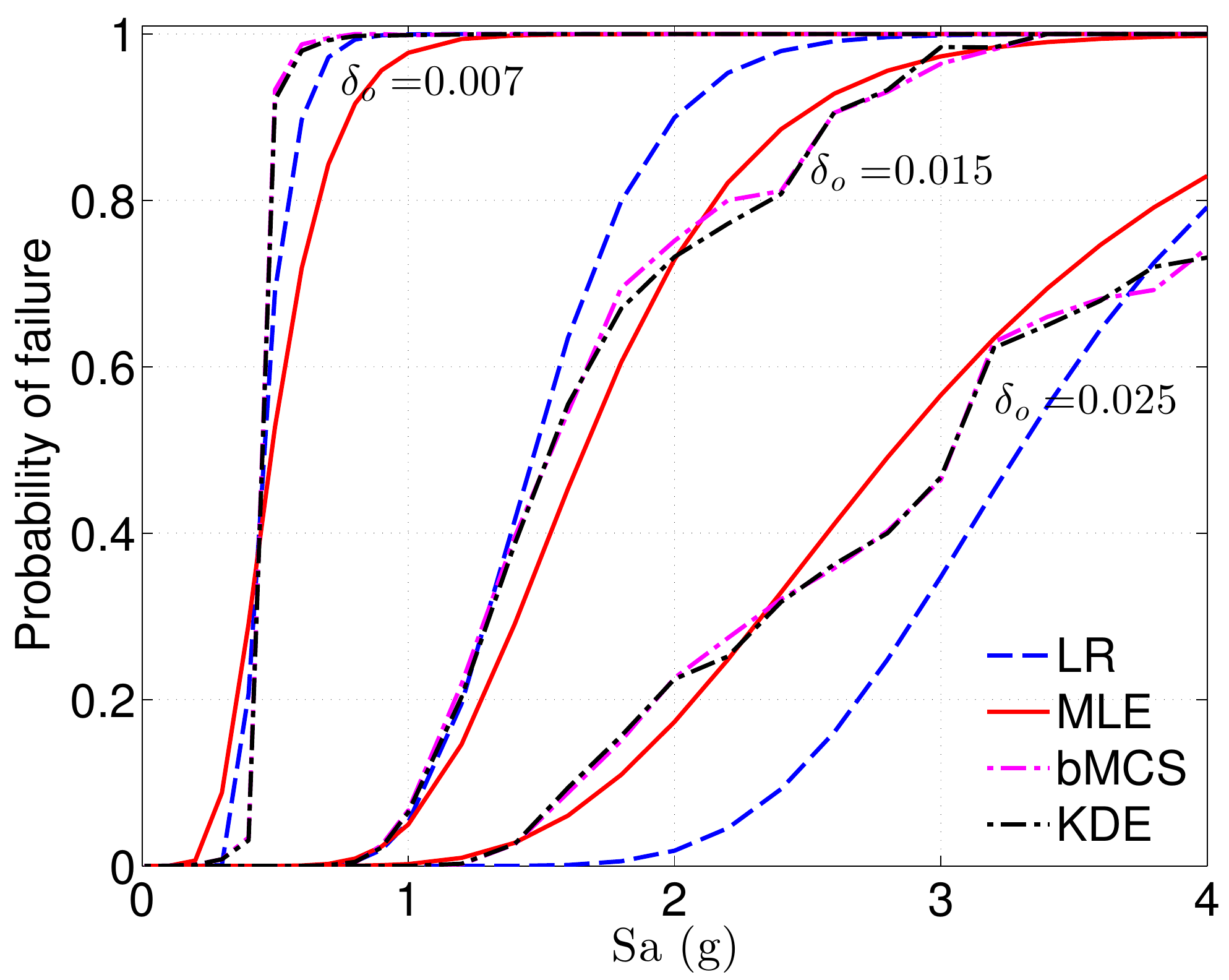}
		\label{fig4.2.2b}
	}
	\caption{Fragility curves with parametric and non-parametric approaches using $PGA$ and $Sa$ as intensity measures (LR: linear regression; MLE: maximum likelihood estimation; bMCS: binned Monte Carlo simulation; KDE: kernel density estimation).}
	\label{fig4.2.2}
\end{figure}

For the case when $PGA$ is considered as $IM$, \figref{fig4.2.2} (left) shows the fragility curves obtained with the MLE- and LR-based lognormal approaches and the bMCS- and KDE-based non-parametric approaches. One first observes a remarkable consistency between the curves obtained with the two non-parametric approaches despite the distinct differences in the underlying algorithms. This validates the accuracy of the proposed methods. For the lower threshold ($\delta_o=0.7\%$), both parametric curves are in good agreement with the non-parametric ones. For the two higher thresholds, the LR-based lognormal curves exhibit significant deviations from the non-parametric ones leading to an overestimation of the failure probabilities. Note that for $\delta_o=1.5\%$ and $\delta_o=2.5\%$, the median $PGA$ (leading to 50\% probability of exceedance) is respectively underestimated by 19\% and 15\% when the LR aproach is used (see \tabref{tab:2}). In contrast, the MLE-based lognormal curves are in a fair agreement with their non-parametric counterparts with the largest discrepancies observed for the highest threshold $\delta_o = 2.5\%$.

\figref{fig4.2.2} (right) shows the resulting fragility curves when $Sa$ is considered as $IM$. The non-parametric curves based on bMCS and KDE remain consistent independently of the drift threshold. For $\delta_o = 0.7\%$, the fragility curves are steep, which is due to the strong correlation between $Sa$ and $\Delta$ when the structure behaves linearly. For this threshold, the LR-based curve is closer to the non-parametric curves than the MLE-based one. For the two larger thresholds, the MLE-based curves are fairly accurate, whereas the LR-based curves exhibit significant deviations from their non-parametric counterparts. In particular, the LR-based curves overestimate the failure probabilities for $\delta_o=1.5\%$ and underestimate the failure probabilities for $\delta_o=2.5\%$. Note that for $\delta_o=1.5\%$, the median $Sa$ is underestimated by 4\%, whereas for $\delta_o=1.5\%$, the median $Sa$ is overestimated by 8\% when the LR aproach is used (see \tabref{tab:2}).

\begin{figure}[!ht]
	\centering
	\subfigure[$PGA = 0.5 ~g$]
	{
		\includegraphics[width=0.47\textwidth]{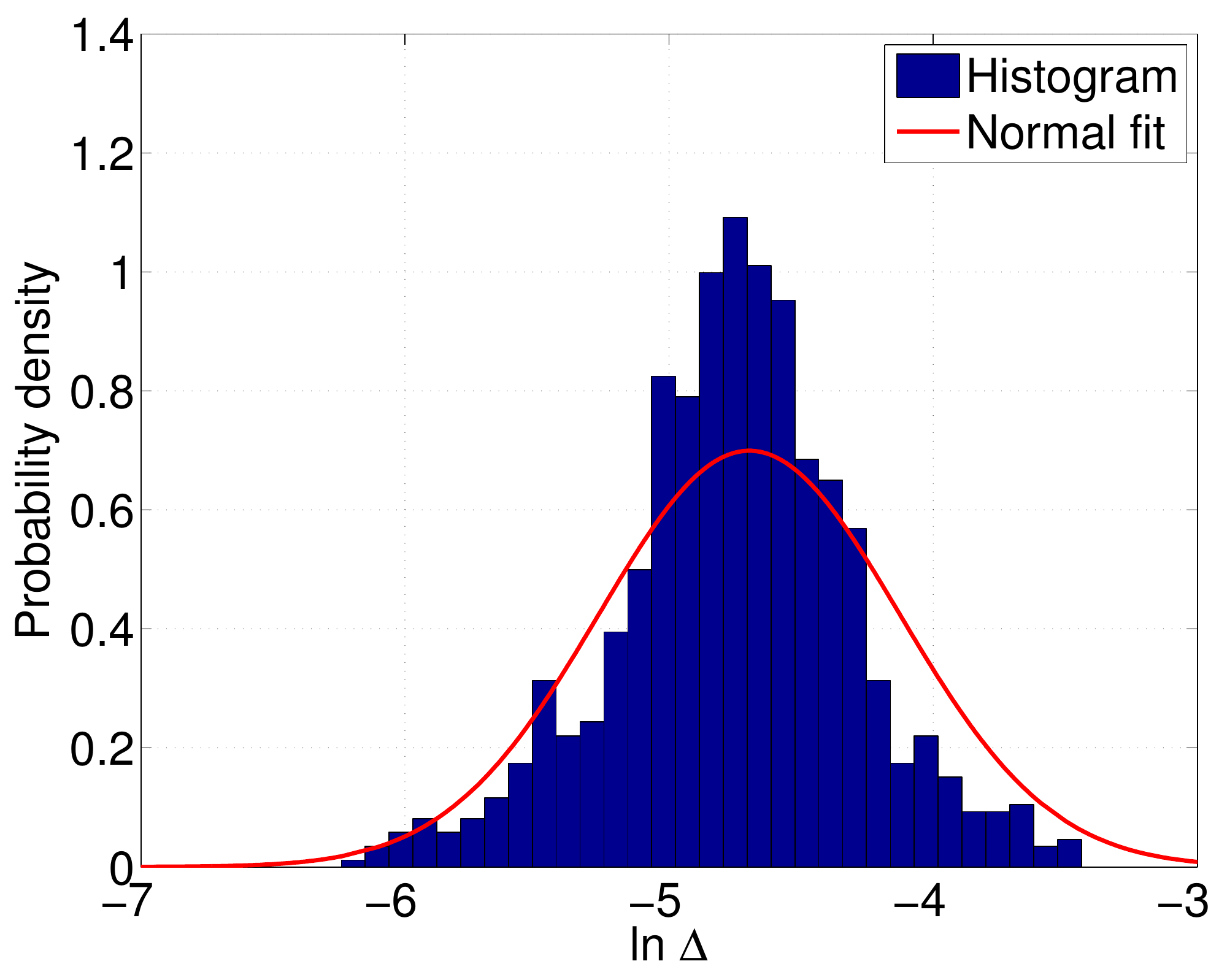}
	}
	\subfigure[$PGA = 1.5 ~g$]
	{
		\includegraphics[width=0.47\textwidth]{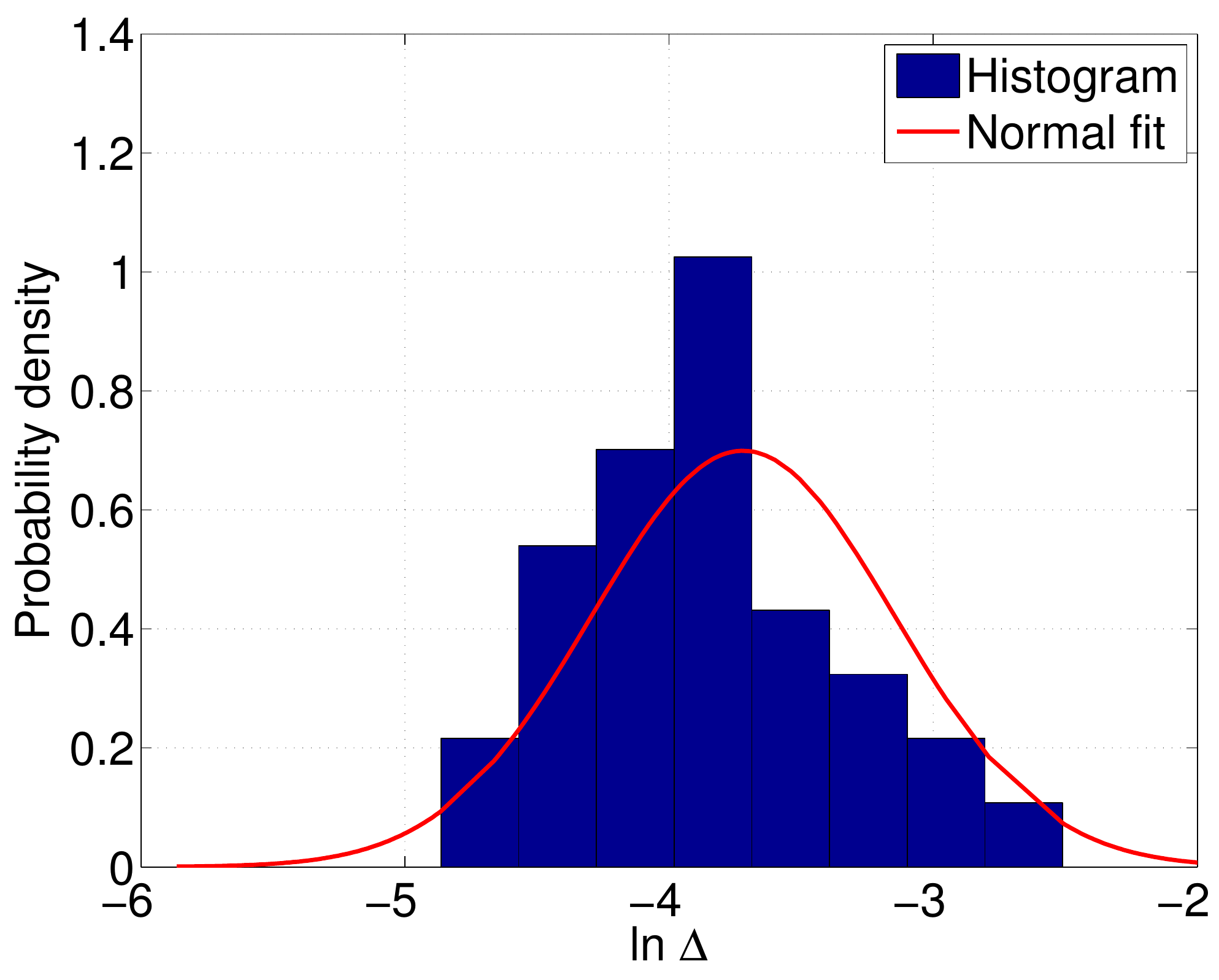}
	}
	\subfigure[$Sa = 0.5 ~g$]
	{
		\includegraphics[width=0.47\textwidth]{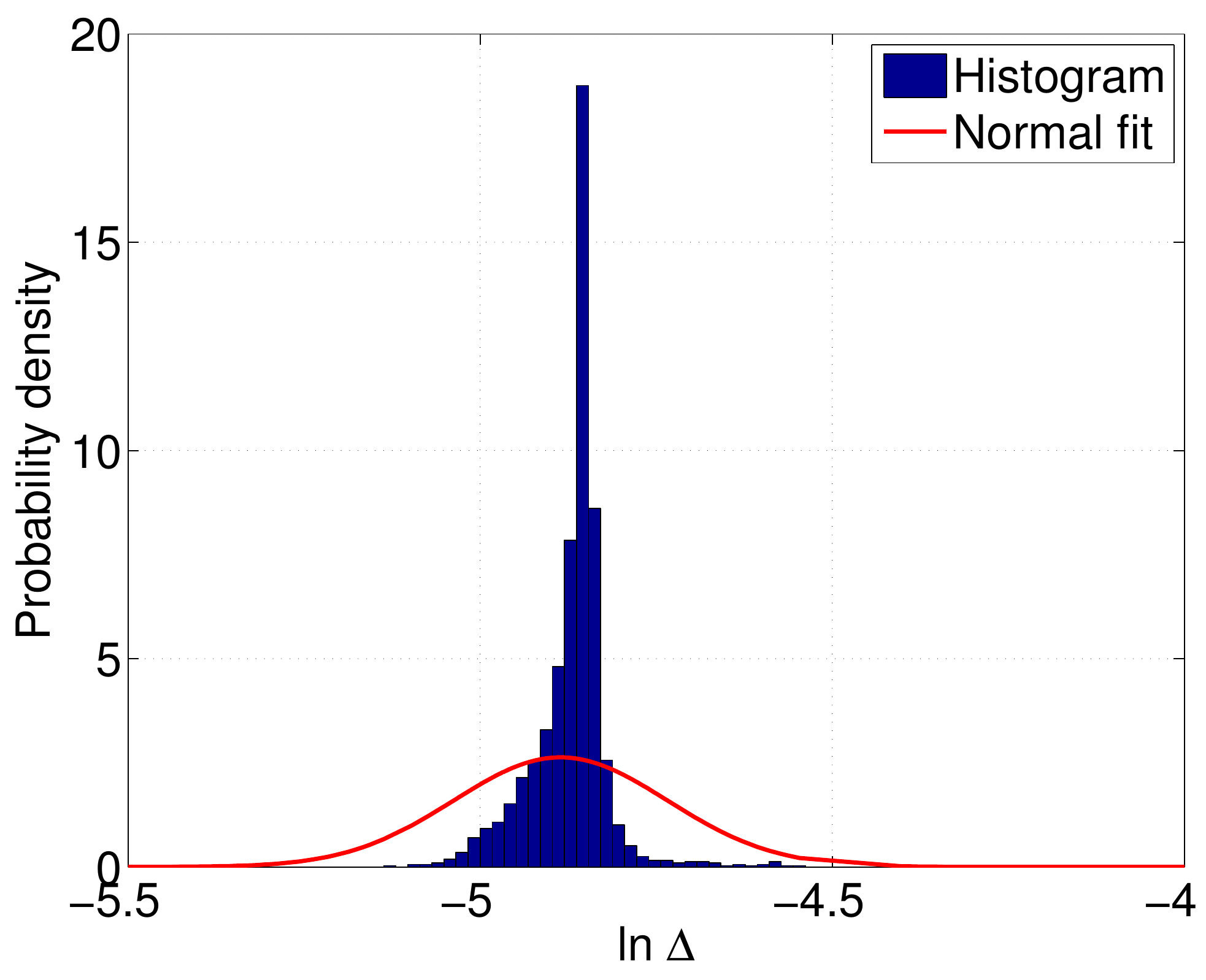}
	}
	\subfigure[$Sa = 1.5 ~g$]
	{
		\includegraphics[width=0.47\textwidth]{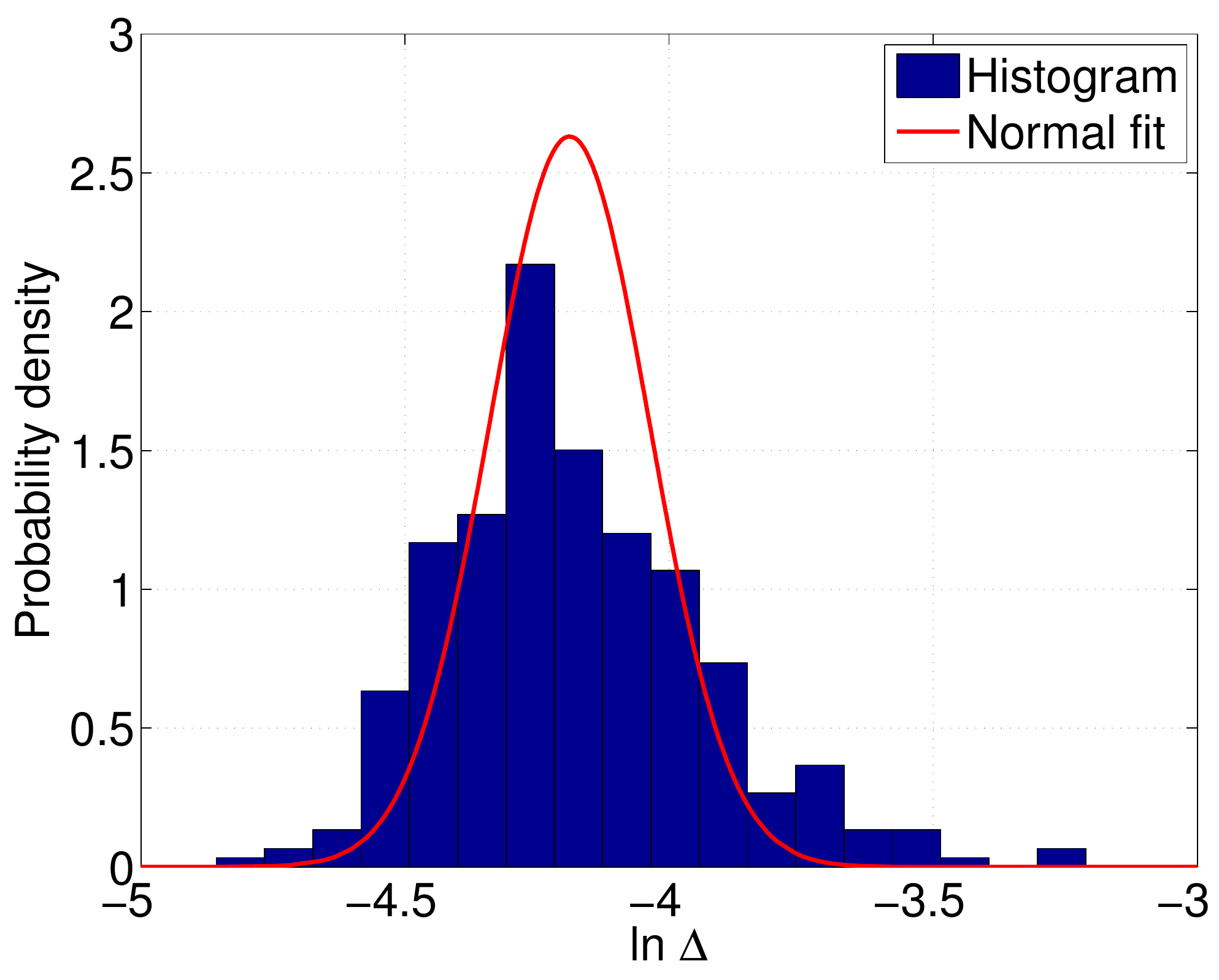}
	}
	\caption{Histograms and fitted normal distributions for $\ln \Delta$ at two levels of $PGA$ and $Sa$.}
	\label{fig4.2.3}
\end{figure}

Summarizing the above results, the MLE-based lognormal approach yields fragility curves that are overall close to the non-parametric ones; however, it smooths out some details of the curves that can be obtained with the non-parametric approaches. On the contrary, the LR-based lognormal curves can be highly inaccurate. As noted in Section \ref{sec2.1.2}, the LR approach assumes that the residuals of the fitted model in the log-scale (\eqrefe{eq8}) follow a normal distribution with a constant standard deviation independently of the $IM$ level. \figref{fig4.2.3} shows histograms of $\ln \Delta$ at two example levels of $PGA$ and $Sa$ together with the fitted normal distributions according to \eqrefe{eq8}. The responses $\Delta$ at each $IM$ level are obtained consistently with the bMCS approach. Obviously, the assumption of a normal distribution is not valid, which is more pronounced when $Sa$ is used as $IM$. This explains the inaccuracy of the LR-based fragility curves for both types of $IM$, despite the relatively high coefficients of determination of the fitted models in the case of $Sa$.

\subsection{Estimation of epistemic uncertainty by bootstrap resampling}

In the following, we use the bootstrap resampling technique (see Section \ref{sec3.4}) to investigate the epistemic uncertainty in the fragility curves estimated with the proposed non-parametric approaches.

\begin{figure}[!ht]
	\centering
	\subfigure
	[Binned Monte Carlo simulation (PGA)]
	{
		\includegraphics[width=0.47\textwidth]{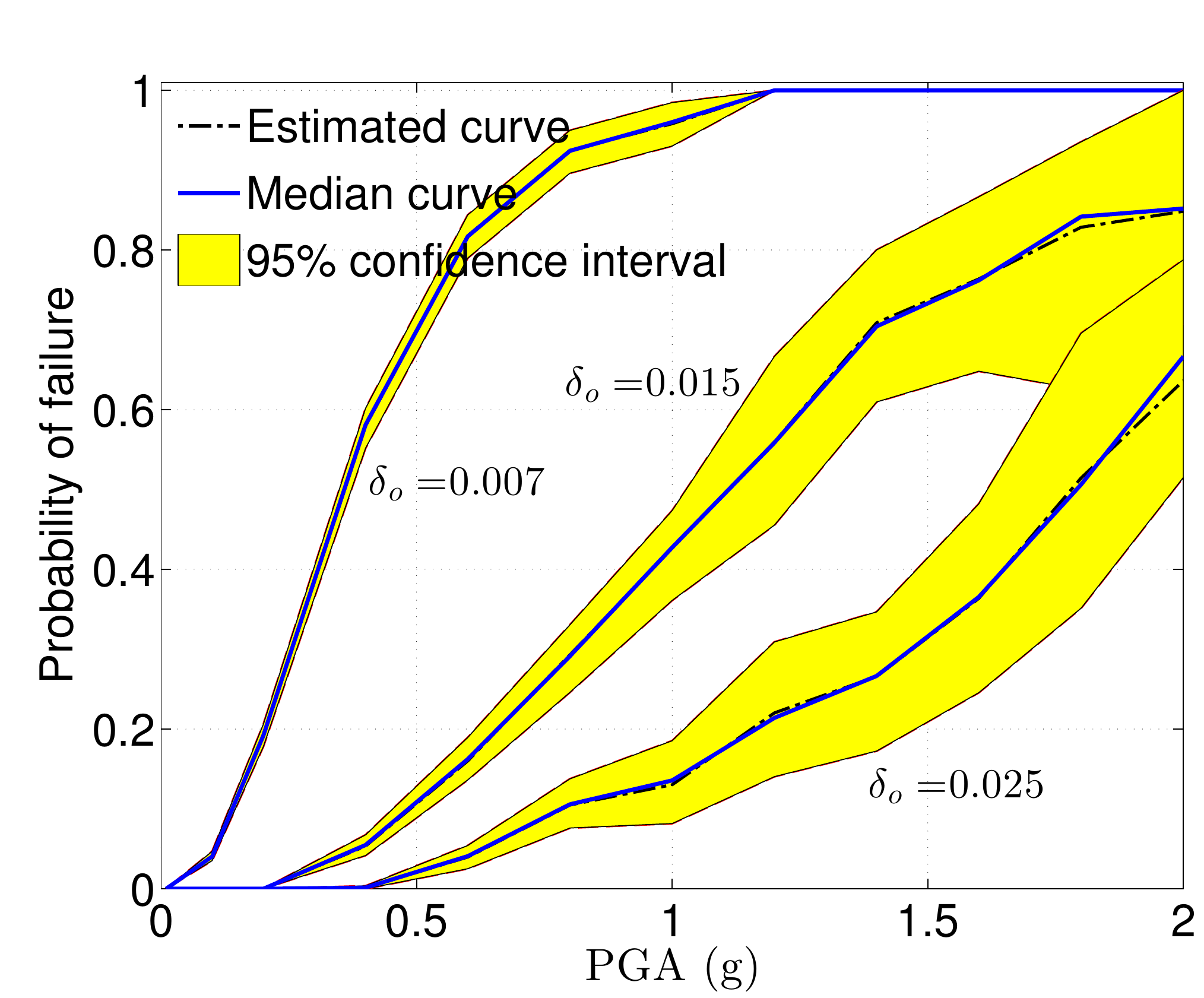}
	}
	\subfigure
	[Kernel density estimation (PGA) ]
	{
		\includegraphics[width=0.47\textwidth]{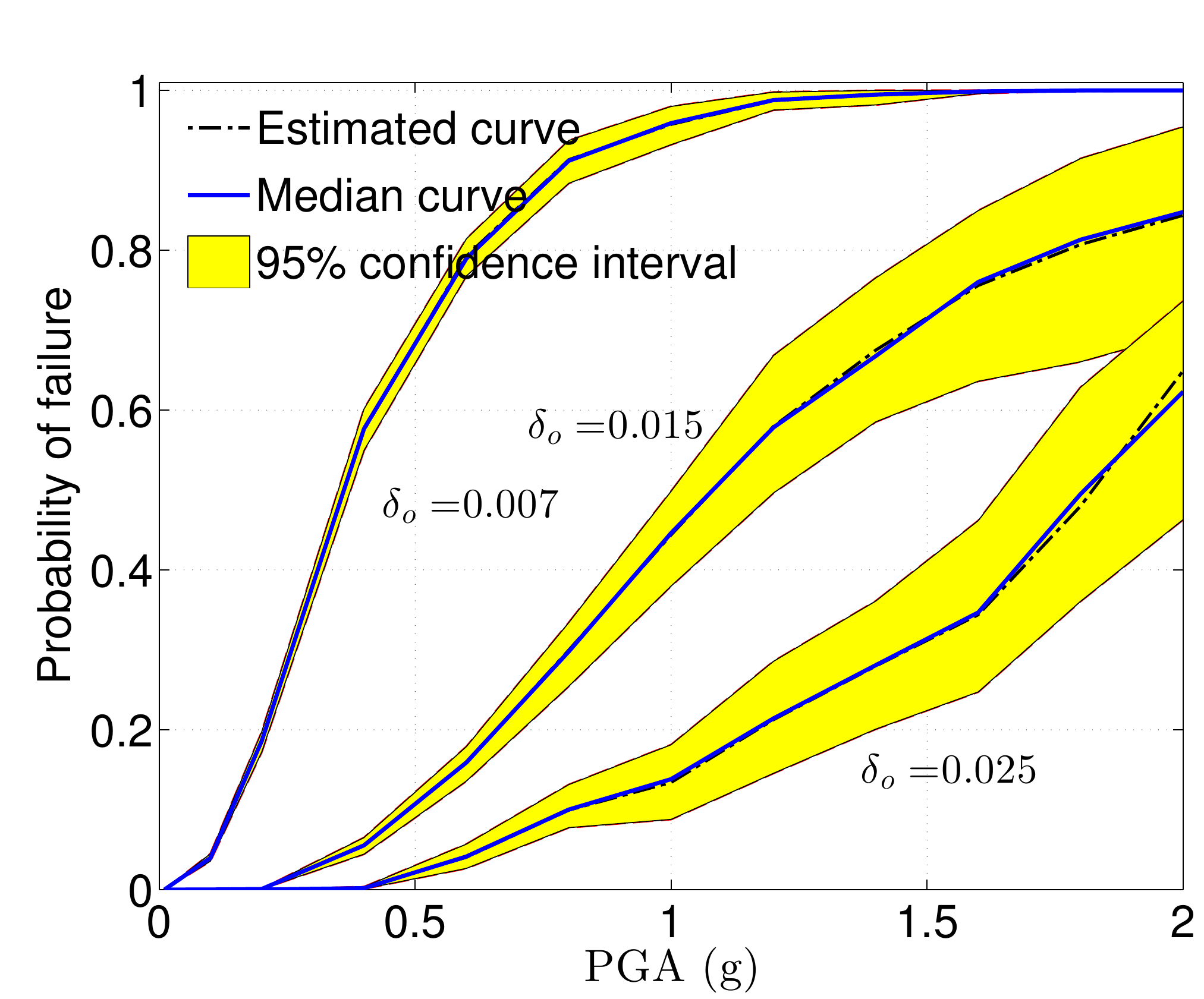}
	}
	\subfigure
	[Binned Monte Carlo simulation (Sa)]
	{
		\includegraphics[width=0.47\textwidth]{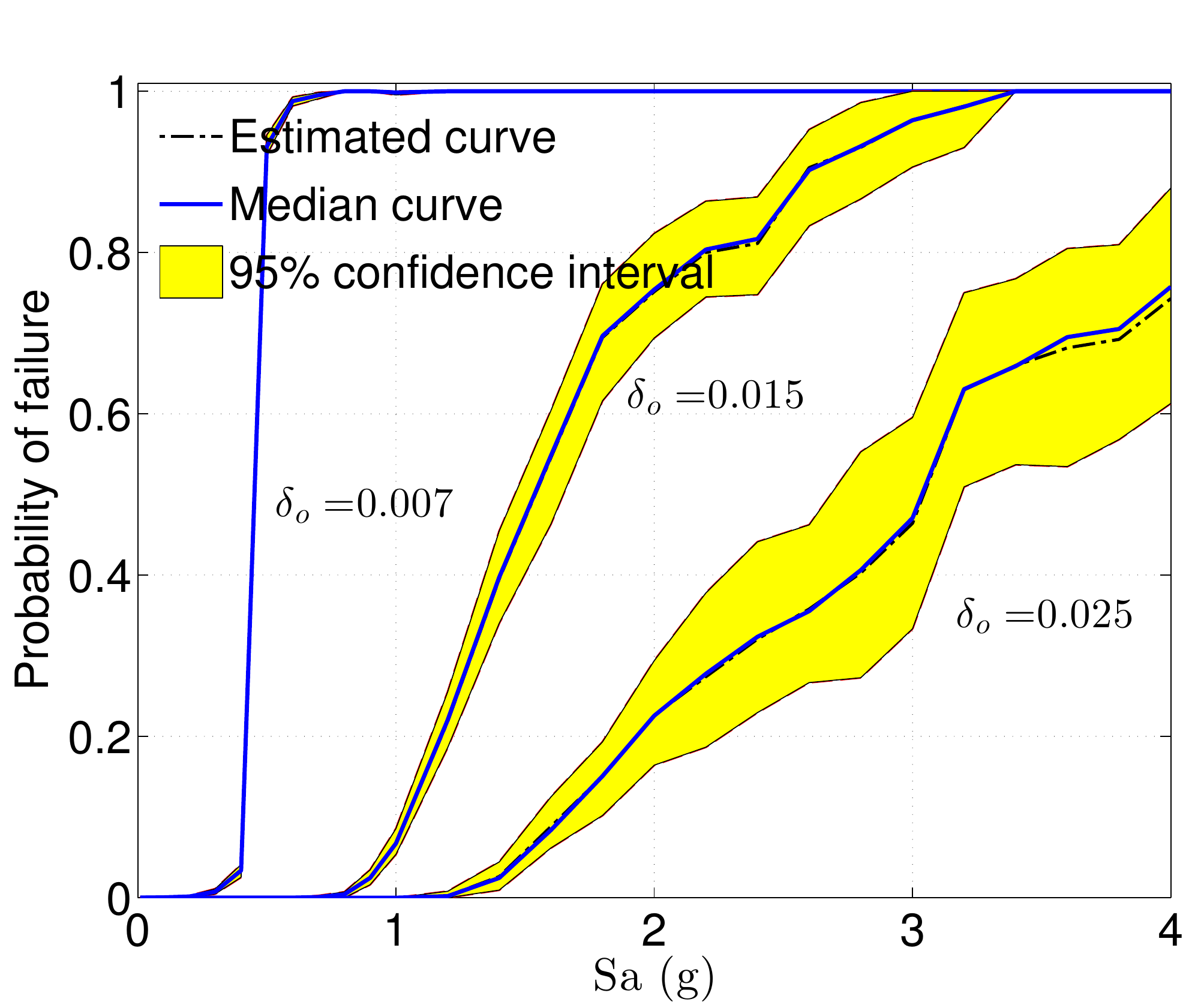}
	}
	\subfigure
	[Kernel density estimation (Sa)]
	{
		\includegraphics[width=0.47\textwidth]{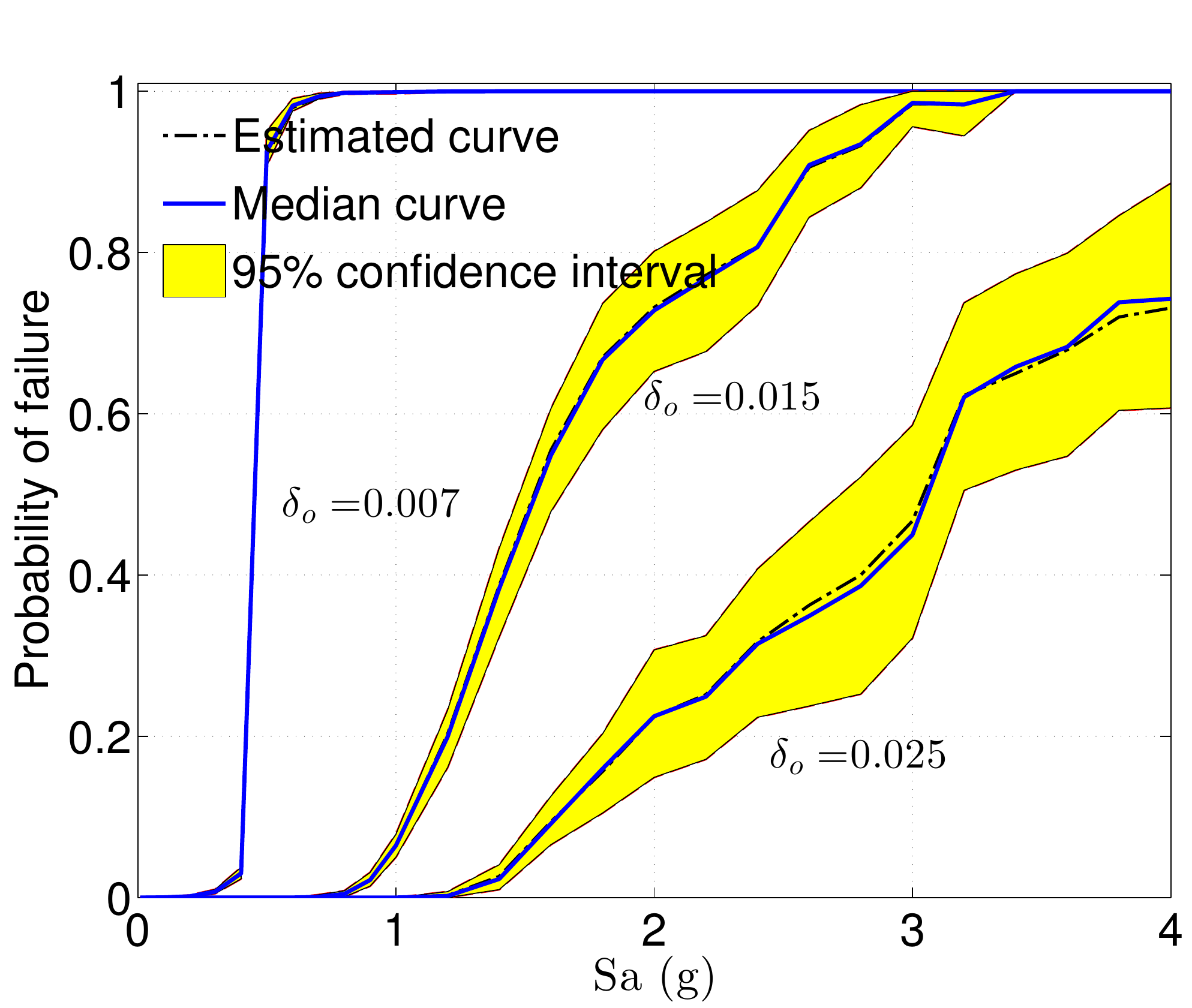}
	}
	\caption{Estimated and mean bootstrap fragility curves and 95\% confidence intervals for the binned Monte Carlo simulation and the kernel density estimation approaches.}
	\label{fig4.3.4}
\end{figure}

We examine the stability of the estimated curves by comparing those with the bootstrap medians, and the variability in the estimation by computing bootstrap confidence intervals. For the two considered $IM$s and the three drift thresholds of interest, \figref{fig4.3.4} shows the median bMCS- and KDE-based fragility curves and the 95\% confidence intervals obtained by bootstrap resampling with 100 replications together with the respective estimated curves (also shown in \figref{fig4.2.2}). \figref{fig4.3.4} clearly shows that both the bMCS-based and the KDE-based median fragility curves obtained with the bootstrap method do not differ from the curves estimated with the original set of observations. This shows the stability of the proposed approaches. For a specified $IM$ and drift limit, the confidence intervals of the bMCS- and KDE-based curves have similar widths. The interval widths tend to increase with increasing drift limit and increasing $IM$ value. 

In order to quantify the effects of epistemic uncertainty, one can estimate the variability of the median $IM$, \ie the $IM$ value leading to 50\% probability of exceedance. Assuming that the median $IM$ ($PGA$ or $Sa$) follows a lognormal distribution \cite{Choun2010}, the median $IM$ is determined for each bootstrap curve and the log-standard deviation of the distribution of the median is computed. Table~\ref{tab:3} lists the log-standard deviations of the median $IM$ values for the same cases as in \figref{fig4.3.4}. These results demonstrate that epistemic uncertainty is increasing with increasing threshold $\delta_o$.  In all cases, the log-standard deviations are relatively small indicating a low level of epistemic uncertainty, which is due to the large number of transient analyses ($N=20,000)$ considered in this study. Although use of such large sets of ground motions is not typical in practice, it is useful for the refined analysis presented here.

\begin{table}[!ht]
	\caption{Log-standard deviation of median $IM$.}
	\centering
	\begin{tabular}{cccccc}
		\toprule
		$\delta_o$ & Approach & $PGA$ & $Sa$  \\
		\midrule  
		\multirow{2}{*}{0.7\%}  
		& bMCS & $0.0003\,g$ &  $0.005\,g$  \\
		& KDE & $0.0005\,g$ & $0.005\,g$  \\
		
		\midrule  
		\multirow{2}{*}{1.5\%}  
		& bMCS & $0.037\,g$ &  $0.054\,g$  \\
		& KDE & $0.037\,g$ &  $0.050\,g$   \\
		
		\midrule  
		\multirow{2}{*}{2.5\%}  
		& bMCS & $0.114\,g$ &  $0.090\,g$  \\
		& KDE & $0.120\,g$ &  $0.080\,g$   \\
		
		\bottomrule
	\end{tabular}
	\label{tab:3}
\end{table}

\section{Concrete column subject to recorded ground motions}

To demonstrate the comparison between the lognormal and the non-parametric approaches for the case when fragility curves are based on recorded ground motions, we herein briefly summarize a case study by the authors originally presented in \cite{MaiEurodyn2014}.

\begin{figure}[!ht]
	\centering
	\begin{subfigure}
		{
			\includegraphics[width=0.55\textwidth]{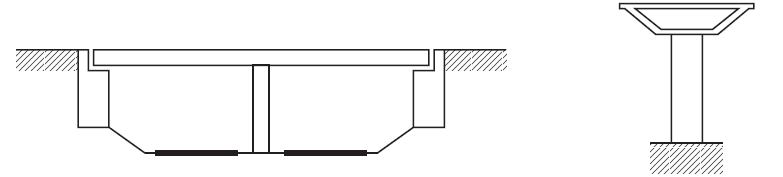}
		}
	\end{subfigure}
	\begin{subfigure}
		{
			\includegraphics[width=0.47\textwidth]{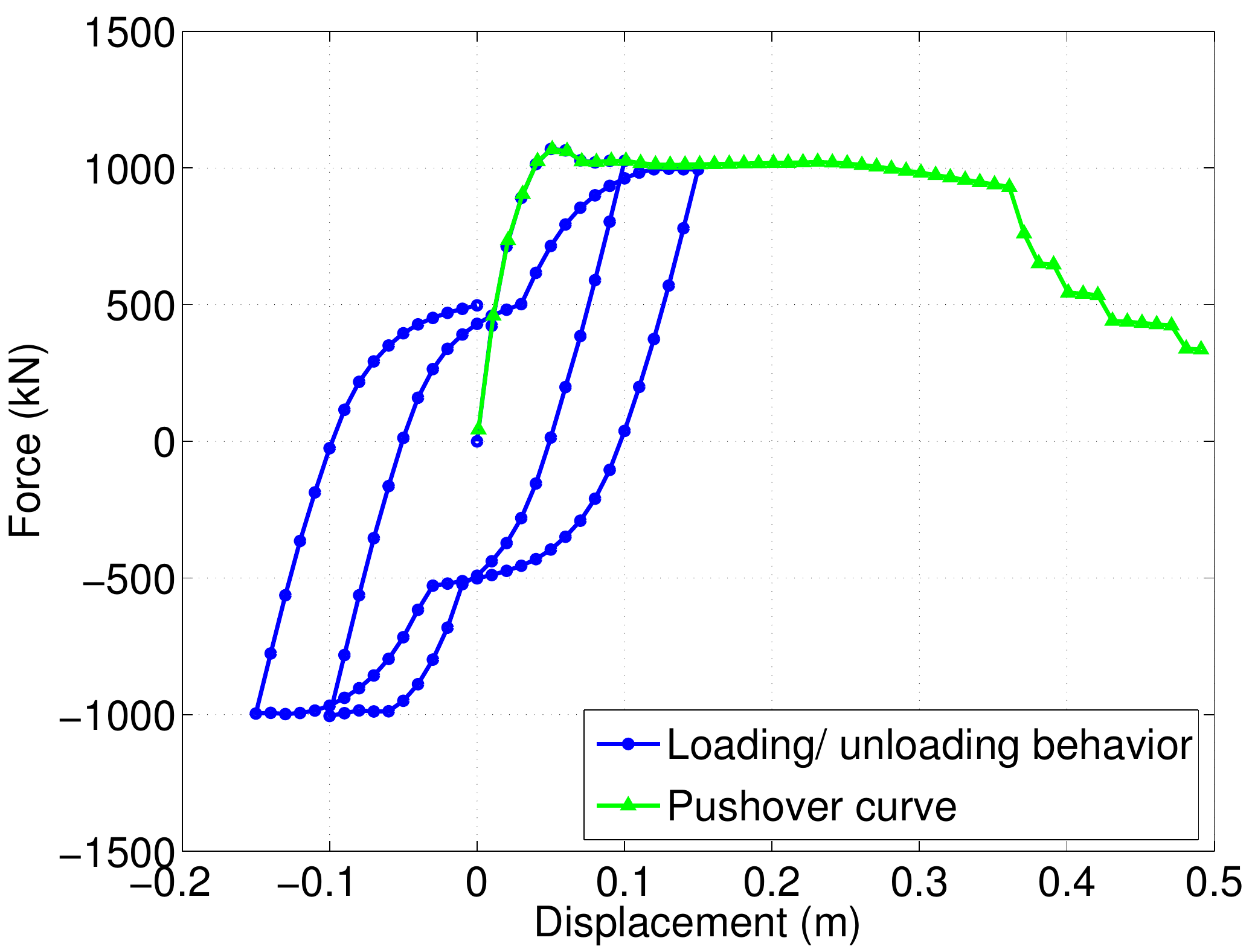}
		}
	\end{subfigure}
	\caption{Bridge configuration \cite{Mackie2003} and hysteretic behavior.}
	\label{fig5.1}
\end{figure}

In this study, we estimate the fragility of a reinforced concrete column with a uniform circular cross-section, representing a pier of a typical California highway overpass bridge \cite{Mackie2003} (see \figref{fig5.1}). The column is modelled in the finite element code OpenSees as a fiberized nonlinear beam-column element. For details on the modelling of the concrete material and the steel reinforcement, the reader is referred to \cite{MaiEurodyn2014}. The loading-unloading behavior and the pushover curve of the column are shown in \figref{fig5.1}. Three-dimensional time-history analyses of the bridge column are conducted for $N=531$ earthquake records (each comprising three orthogonal component accelerograms). These records are obtained from the PEER strong motion database and cover a wide range of source-to-site distances and earthquake moment magnitudes \cite{Mackie2003}. The developed fragility curves represent the probability of the maximal drift ratio $\Delta$ in the transverse direction exceeding specified thresholds $\delta_o$ as a function of the peak ground acceleration $PGA$ or the pseudo-spectral acceleration $Psa$ corresponding to the first transverse mode ($T_1 = 0.535~s$). The considered drift ratio thresholds, shown in Table~\ref{tab5.1}, are recommended for the operational and life safety levels by two different sources \cite{Marsh2013,Lu2005}.

\begin{table}[!ht]
	\caption{Bridge performance and respective drift-ratio threshold.}
	\centering
	\begin{tabular}{llccc}
		\hline
		Reference &Level & Description & Damage & Drift ratio $\delta_o$  \\
		\hline
		\cite{Marsh2013}& II & Operational & Minor &  0.01 \\
		\cite{Marsh2013} & III & Life safety & Moderate  &  0.03  \\
		\cite{Lu2005} & II & Operational & Minor  &  0.005  \\
		\cite{Lu2005} & III & Life safety & Moderate  &  0.015  \\
		\hline
	\end{tabular}
	\label{tab5.1}
\end{table}

\begin{figure}[!ht]
	\centering
	\subfigure
	{
		\includegraphics[width=0.47\textwidth]{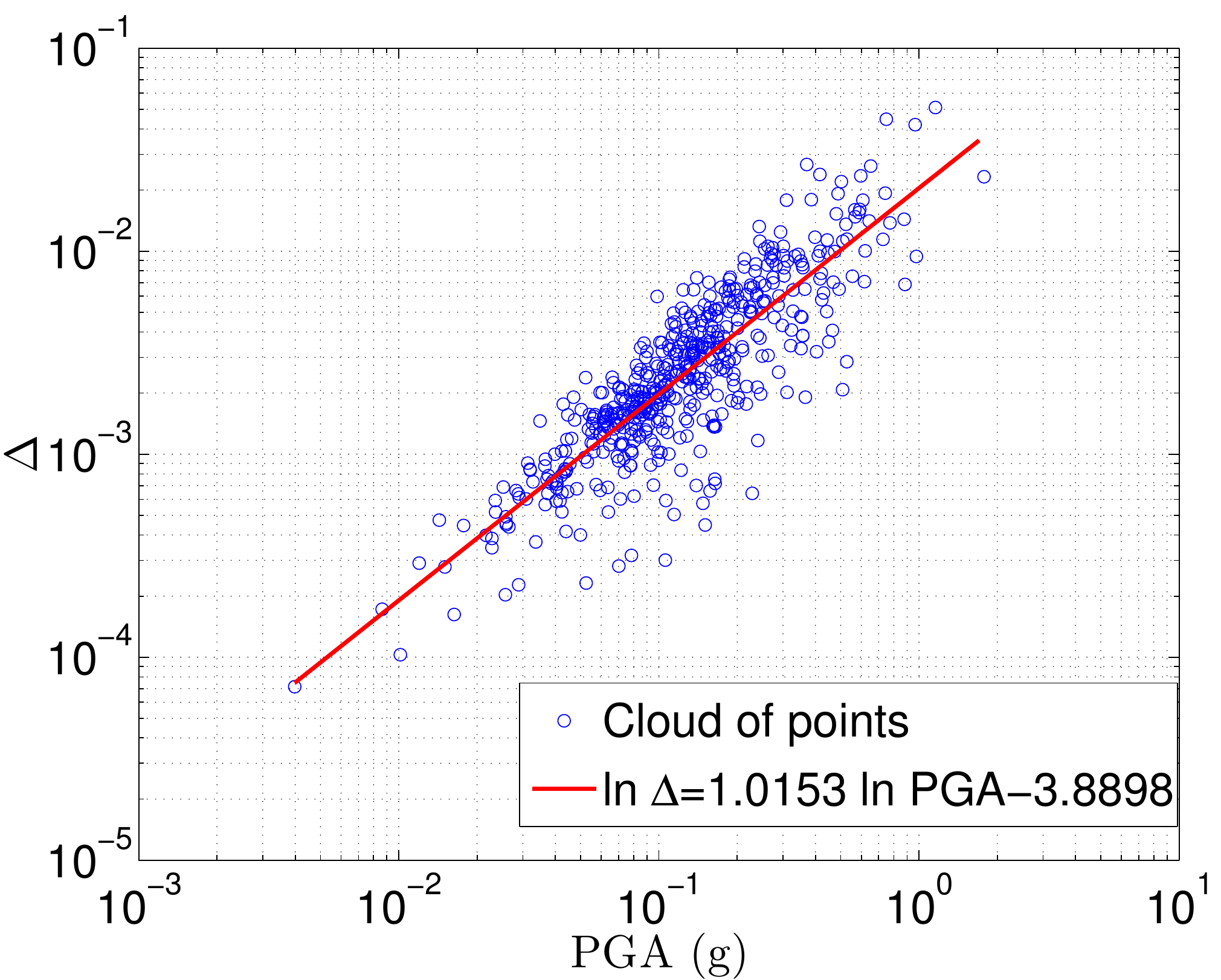}
	}
	\subfigure
	{
		\includegraphics[width=0.47\textwidth]{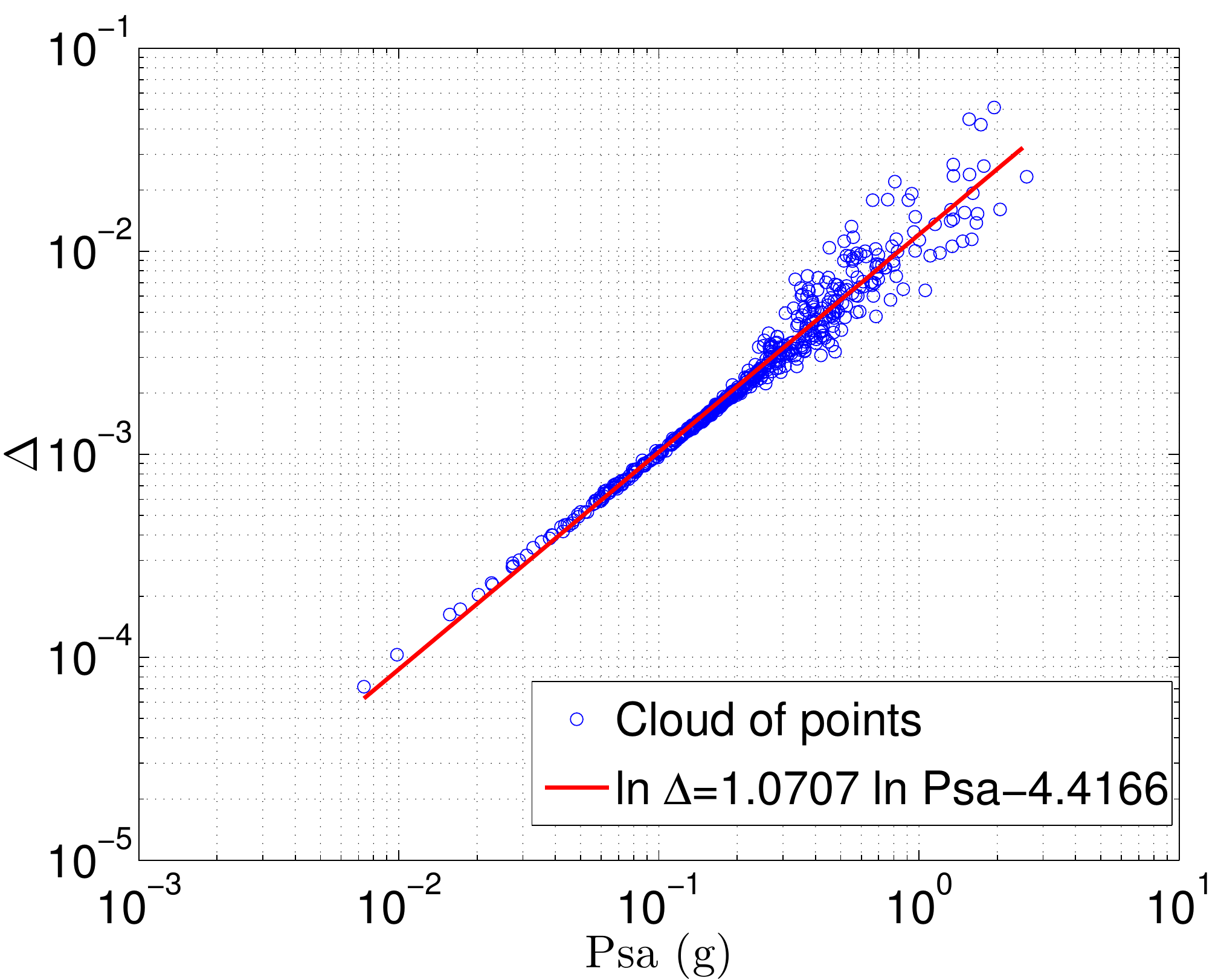}
	}
	\caption{Paired data $\acc{\prt{IM_i,\Delta_i}, i=1 \enu N}$ and fitted models in log-scale (the units of the variables in the fitted models are the same as in the axes of the graphs).}
	\label{fig5.2}
\end{figure}

The fragility curves are established with the MLE-based and LR-based lognormal approaches and the KDE-based non-parametric approach. Due to the relatively small number of data, the bMCS method is not considered herein. \figref{fig5.2} depicts the clouds of points $(IM_i, \, \Delta_i)$ in the logarithmic scale for the two $IM$s together with the linear fitted models. The coefficients of determination of the latter are $R^2=0.729$ for the case of $PGA$ and $R^2=0.963$ for the case of $Psa$. Note that for small values of $Psa$ ($Psa<$0.2~g) a linear function provides a perfect fit, which is due to the fact that in this range of $Psa$, the column behavior can be represented by a linear single-degree-of-freedom model. Further details on the parameters of the different fragility functions can be found in \cite{MaiEurodyn2014}.

\begin{figure}[!ht]
	\centering
	\subfigure
	{
		\includegraphics[width=0.47\textwidth]{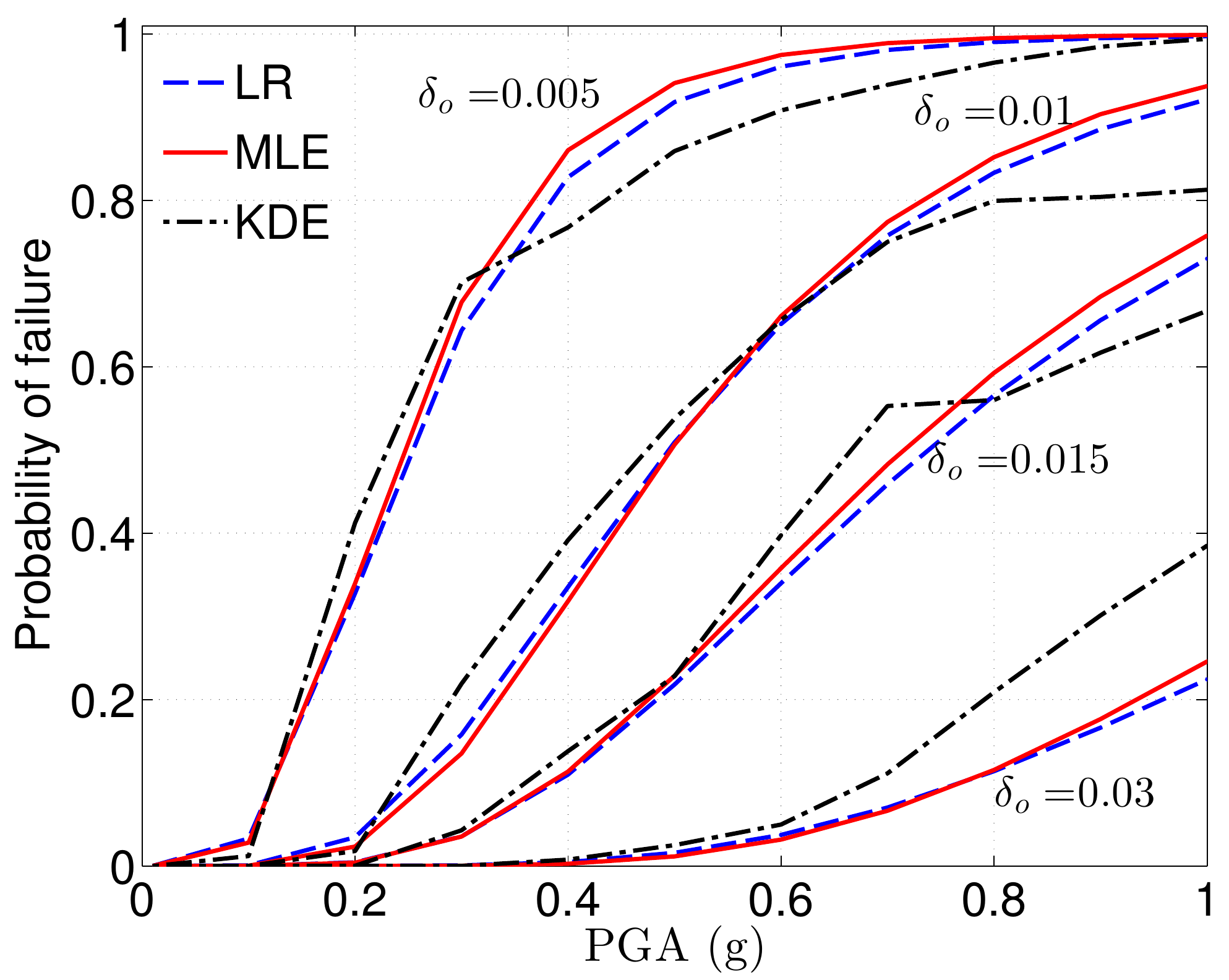}
	}
	\subfigure
	{
		\includegraphics[width=0.47\textwidth]{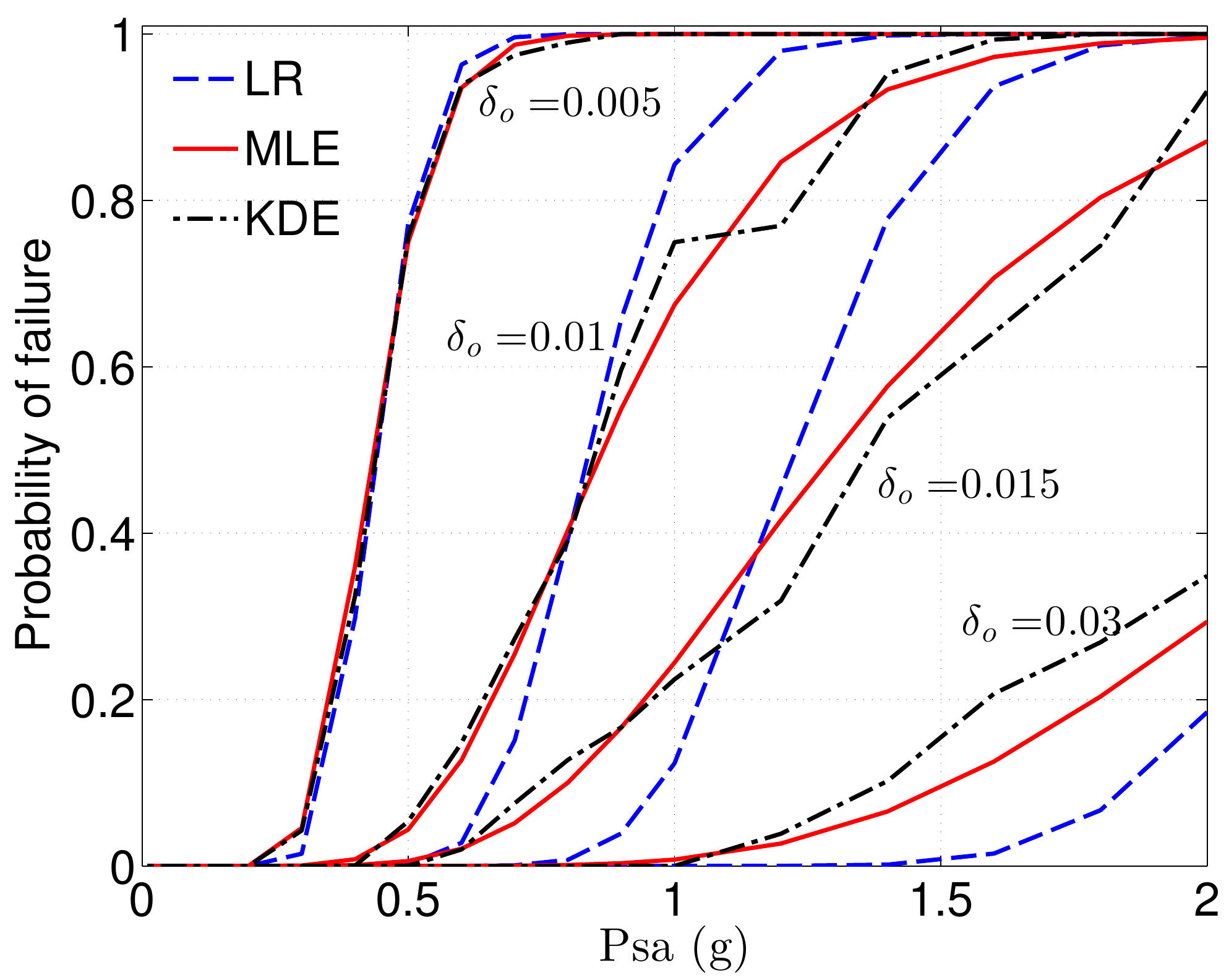}
	}
	\caption{Fragility curves with parametric and non-parametric approaches using $PGA$ and $Psa$ as intensity measures (LR: linear regression; MLE: maximum likelihood estimation; KDE: kernel density estimation.}
	\label{fig5.3}
\end{figure}

\figref{fig5.3} depicts the obtained fragility curves for the two types of $IM$ and the four drift-ratio thresholds of interest. When $PGA$ is used as $IM$, the curves obtained with the two lognormal approaches are close to each other for all thresholds, but exhibit deviations from the non-parametric curves, which tend to be larger for higher PGA levels and larger drift limits. When $Psa$ is used as $IM$, the MLE-based curves are in a fair agreement with the KDE-based ones; however, the former smooth out some details that can be obtained with the non-parametric approach. In contrast, the LR-based curves are inaccurate for all but the smaller drift threshold. Overall, the LR-based curves exhibit larger deviations from the non-parametric ones for $Psa$ than for $PGA$ as $IM$, although the $R^2$ coefficient of the linear fit is higher for $Psa$. This can be explained by the fact that the assumption of \textit{homoscedastic errors}, inherent in \eqrefe{eq8}, is not valid for the specific data set $(Psa,\Delta)$, as one can observe in \figref{fig5.2}.

\section{Discussion}

Using the non-parametric fragility curves as reference, the accuracy of the lognormal curves is found to depend on the method used to estimate the parameters of the underlying CDF, the considered $IM$ and the drift threshold of interest. In most cases, the MLE-based curves are fairly close to the non-parametric ones, whereas the LR-based curves exhibit significant deviations. The lognormal curves tend to deviate more from the non-parametric ones for larger drift limits. Considering both case studies, the MLE-based curves are more accurate for a structure-specific $IM$ ($Sa$, $Psa$) than for $PGA$. Different $IM$s have been recommended in the literature for structures of different type, size and material \cite{Padgett2008a,Jankovic2004}. Accordingly, the accuracy of the lognormal fragility curves may depend on those factors as well. Possible dependence of the accuracy of the lognormal curves on the considered response quantity needs to be investigated as well.

As noted in Section \ref{sec3.2}, the bMCS approach bears similarities with the so-called stripe analysis \cite{Shome1998,Bazzurro1998,Jalayer2009}. A comparison between the stripe and cloud analyses, where the latter corresponds to the LR-based lognormal approach \cite{Cornell2002,Baker2007,Celik2010,Jalayer2014}, was carried out by Celik and Ellingwood \cite{Celik2010}. In the mentioned study, concrete structures were subject to 40 synthetic ground motions. Differences in the response statistics obtained with the two methods for three $IM$ levels were found insignificant and hence, use of the cloud analysis was justified. In contrast, Baker \cite{Baker2007} showed that cloud analysis can significantly underestimate the mean annual rate of exceeding a large maximum interstory drift. Karamlou and Bocchini \cite{Karamlou2015} recently conducted large-scale simulations on bridge structures in order to investigate the underlying assumptions of the cloud analysis. Their results showed that, in general, the conditional distribution of a demand parameter for a given $IM$ level is not lognormal. In addition, it was found that the assumptions of a linear function for the probabilistic seismic demand model in the log-scale (power function in the normal scale) and of constant dispersion of the respective errors can lead to significant errors in fragility analysis. These findings are consistent with our results shown in \figref{fig4.2.3}. The limitations of the LR-based approach have also been mentioned by Jalayer et al. \cite{Jalayer2014}.

Based on the results of our case studies and the above discussion, we recommend the use of the MLE approach if fragility curves are developed in a parametric manner. The superiority of the MLE over the LR approach relies on the fact that the former avoids the assumptions of the linear model and the homoscedasticity of the errors that are inherent in the latter. However, when a detailed description of the fragility function is important, a non-parametric approach should be used. The bMCS method requires a large number of data, which can be typically obtained by use of synthetic motions; note that due to the current computer capacities and the use of distributed computing, large-scale simulations are becoming increasingly popular among both researchers and practitioners. On the other hand, the KDE approach can be employed even with a limited number of recorded motions at hand, as shown in our second case study. We again emphasize that the two non-parametric approaches lead to almost identical curves in the case when they could be applied independently with the same (large) dataset.

\section{Conclusions}

Seismic demand fragility evaluation is one of the basic elements in the framework of performance-based earthquake engineering (PBEE). At present, the classical lognormal approach is widely used to establish such fragility curves mainly due to the fact that the lognormality assumption makes seismic risk analysis more tractable. The approach consists in assigning the shape of a lognormal cumulative distribution function to the fragility curves. However, the validity of this assumption remains an open question.

In this paper, we introduce two non-parametric approaches in order to examine the validity of the classical lognormal approach, namely the binned Monte Carlo simulation and the kernel density estimation. The former computes the crude Monte Carlo estimators for small subsets of ground motions with similar values of a selected intensity measure, while the latter estimates the conditional probability density function of the structural response given the ground motion intensity measures using the kernel density estimation technique. The proposed approaches can be used to compute fragility curves when the actual shape of these curves is not known as well as to validate or calibrate parametric fragility curves. Herein, the two non-parametric approaches are confronted to the classical lognormal approach in two case studies, considering synthetic and recorded ground motions.

In the case studies, the fragility curves are established for various drift thresholds and different types of the ground motion intensity measure, namely the peak ground acceleration ($PGA$), and the structure-specific spectral acceleration ($Sa$) and pseudo-spectral acceleration ($Psa$). The two non-parametric curves are always consistent, which proves the validity of the proposed techniques. Accordingly, the non-parametric curves are used as reference to assess the accuracy of the lognormal curves. The parameters of the latter are estimated with two approaches, namely by maximum likelihood estimation and by assuming a linear probabilistic seismic demand model in the log-scale. The maximum likelihood estimation approach is found to approximate fairly well the reference curves in most cases, especially when a structure-specific intensity measure is used; however, it smooths out some details that can be obtained with the non-parametric approaches. In contrast, the assumption of a linear demand model in the log-scale is found overall inaccurate. When integrated in the PBEE framework, inaccuracy in fragility estimation may induce errors in the probabilistic consequence estimates that serve as decision variables for risk mitigation actions.
The bootstrap resampling technique is employed to assess effects of epistemic uncertainty in the non-parametric fragility curves. Results from bootstrap analysis validate the stability of the fragility estimates with the proposed non-parametric methods.

Recently, fragility surfaces have emerged as an innovative way to represent the vulnerability of a system \cite{Seyedi2010}; these represent the failure probability conditional on two intensity measures of the earthquake motions. The computation of these surfaces is not straightforward and requires a large computational effort. The present study opens new paths for establishing the fragility surfaces:
similarly to the case of fragility curves, one can use kernel density estimation to obtain fragility surfaces that are free of the lognormality assumption and consistent with the surfaces obtained by Monte Carlo simulation.

We note that the computational cost of the two proposed approaches is significant when they are based on large Monte Carlo samples. In order to reduce this cost, alternative approaches may be envisaged. 
Polynomial chaos (PC) expansions \cite{Ghanembook2003,BlatmanJCP2011} 
appear as a promising tool. Based on a smaller sample set (typically a few hundreds of finite element runs), PC expansion provides a polynomial approximation that surrogates the structural response. The feasibility of post-processing PC expansions in order to compute fragility curves has been shown in~\cite{SudretGuyonnet2011,SudretMaiIcossar2013} in the case a linear structural behavior is assumed. The extension to nonlinear behavior is currently in progress.

We underline that the proposed non-parametric approaches are essentially applicable to other probabilistic models in the PBEE framework, relating decision variables with structural damage and structural damage with structural response.  Once all the non-parametric probabilistic models are available, they can be incorporated in the PBEE framework by means of numerical integration. Then a full seismic risk assessment may be conducted by avoiding potential inaccuracies introduced from simplifying parametric assumptions at any step of the analysis. Optimal high-fidelity computational methods for incorporating non-parametric fragility curves in the PBEE framework will be investigated in the future.


\section*{Acknowledgements}
	The authors are thankful to the two anonymous reviewers for various valuable comments that helped improve the quality of the manuscript. Discussions with Dr. Sanaz Rezaeian, who provided clarifications on the stochastic ground motion model used in this study, are also acknowledged.





\begin{thebibliography}{}
	\providecommand{\natexlab}[1]{#1}
	\providecommand{\url}[1]{\texttt{#1}}
	\providecommand{\href}[2]{#2}
	\providecommand{\path}[1]{#1}
	\providecommand{\eprint}[1]{\href{http://arxiv.org/abs/#1}{\path{#1}}}
	\providecommand{\DOIprefix}{doi:}
	\providecommand{\ArXivprefix}{arXiv:}
	\providecommand{\URLprefix}{URL: }
	\providecommand{\Pubmedprefix}{pmid:}
	\providecommand{\doi}[1]{\href{http://dx.doi.org/#1}{\path{#1}}}
	\providecommand{\Pubmed}[1]{\href{pmid:#1}{\path{#1}}}
	\providecommand{\BIBand}{and}
	\providecommand{\bibinfo}[2]{#2}
	\ifx\xfnm\undefined \def\xfnm[#1]{\unskip,\space#1}\fi
	\bibitem{Porter2003}
	\bibinfo{author}{Porter\xfnm[ K.A.]}.
	\newblock \bibinfo{title}{{An overview of PEER's performance-based earthquake
			engineering methodology}}.
	\newblock In: \bibinfo{booktitle}{Proc. 9th Int. Conf. on Applications of Stat.
		and Prob. in Civil Engineering (ICASP9), San Francisco}.
	\bibinfo{year}{2003}, p. \bibinfo{pages}{6--9}.
	\bibitem{Baker2008a}
	\bibinfo{author}{Baker\xfnm[ J.W.]}, \bibinfo{author}{Cornell\xfnm[ C.A.]}.
	\newblock \bibinfo{title}{{Uncertainty propagation in probabilistic seismic
			loss estimation}}.
	\newblock \bibinfo{journal}{Structural Safety}
	\bibinfo{year}{2008};\bibinfo{volume}{30}(\bibinfo{number}{3}):\bibinfo{pages}{236--252}.
	\bibitem{Gunay2013}
	\bibinfo{author}{G\"{u}nay\xfnm[ S.]}, \bibinfo{author}{Mosalam\xfnm[ K.M.]}.
	\newblock \bibinfo{title}{{PEER performance-based earthquake engineering
			methodology, revisited}}.
	\newblock \bibinfo{journal}{J Earthq Eng}
	\bibinfo{year}{2013};\bibinfo{volume}{17}(\bibinfo{number}{6}):\bibinfo{pages}{829--858}.
	\bibitem{Mackie2005}
	\bibinfo{author}{Mackie\xfnm[ K.]}, \bibinfo{author}{Stojadinovic\xfnm[ B.]}.
	\newblock \bibinfo{title}{{Fragility basis for California highway overpass
			bridge seismic decision making}}.
	\newblock \bibinfo{publisher}{Pacific Earthquake Engineering Research Center,
		College of Engineering, University of California, Berkeley};
	\bibinfo{year}{2005}.
	\bibitem{Ellingwood2009}
	\bibinfo{author}{Ellingwood\xfnm[ B.R.]}, \bibinfo{author}{Kinali\xfnm[ K.]}.
	\newblock \bibinfo{title}{{Quantifying and communicating uncertainty in seismic
			risk assessment}}.
	\newblock \bibinfo{journal}{Structural Safety}
	\bibinfo{year}{2009};\bibinfo{volume}{31}(\bibinfo{number}{2}):\bibinfo{pages}{179--187}.
	\bibitem{Seo2012}
	\bibinfo{author}{Seo\xfnm[ J.]}, \bibinfo{author}{Duenas-Osorio\xfnm[ L.]},
	\bibinfo{author}{Craig\xfnm[ J.I.]}, \bibinfo{author}{Goodno\xfnm[ B.J.]}.
	\newblock \bibinfo{title}{{Metamodel-based regional vulnerability estimate of
			irregular steel moment-frame structures subjected to earthquake events}}.
	\newblock \bibinfo{journal}{Eng Struct}
	\bibinfo{year}{2012};\bibinfo{volume}{45}:\bibinfo{pages}{585--597}.
	\bibitem{Banerjee2007}
	\bibinfo{author}{Banerjee\xfnm[ S.]}, \bibinfo{author}{Shinozuka\xfnm[ M.]}.
	\newblock \bibinfo{title}{{Nonlinear static procedure for seismic vulnerability
			assessment of bridges}}.
	\newblock \bibinfo{journal}{Comput-Aided Civ Inf}
	\bibinfo{year}{2007};\bibinfo{volume}{22}(\bibinfo{number}{4}):\bibinfo{pages}{293--305}.
	\bibitem{Richardson1980}
	\bibinfo{author}{Richardson\xfnm[ J.E.]}, \bibinfo{author}{Bagchi\xfnm[ G.]},
	\bibinfo{author}{Brazee\xfnm[ R.J.]}.
	\newblock \bibinfo{title}{{The seismic safety margins research program of the
			U.S. Nuclear Regulatory Commission}}.
	\newblock \bibinfo{journal}{Nuc Eng Des}
	\bibinfo{year}{1980};\bibinfo{volume}{59}(\bibinfo{number}{1}):\bibinfo{pages}{15--25}.
	\bibitem{Pei2009}
	\bibinfo{author}{Pei\xfnm[ S.]}, \bibinfo{author}{{Van De Lindt}\xfnm[ J.]}.
	\newblock \bibinfo{title}{{Methodology for earthquake-induced loss estimation:
			An application to woodframe buildings}}.
	\newblock \bibinfo{journal}{Structural Safety}
	\bibinfo{year}{2009};\bibinfo{volume}{31}(\bibinfo{number}{1}):\bibinfo{pages}{31--42}.
	\bibitem{Eads2013}
	\bibinfo{author}{Eads\xfnm[ L.]}, \bibinfo{author}{Miranda\xfnm[ E.]},
	\bibinfo{author}{Krawinkler\xfnm[ H.]}, \bibinfo{author}{Lignos\xfnm[ D.G.]}.
	\newblock \bibinfo{title}{{An efficient method for estimating the collapse risk
			of structures in seismic regions}}.
	\newblock \bibinfo{journal}{Earthquake Eng Struct Dyn}
	\bibinfo{year}{2013};\bibinfo{volume}{42}(\bibinfo{number}{1}):\bibinfo{pages}{25--41}.
	\bibitem{Dukes2012}
	\bibinfo{author}{Dukes\xfnm[ J.]}, \bibinfo{author}{DesRoches\xfnm[ R.]},
	\bibinfo{author}{Padgett\xfnm[ J.E.]}.
	\newblock \bibinfo{title}{{Sensitivity study of design parameters used to
			develop bridge specific fragility curves}}.
	\newblock In: \bibinfo{booktitle}{Proc. 15th World Conf. Earthquake Eng.}
	\bibinfo{year}{2012},.
	\bibitem{Guneyisi2008}
	\bibinfo{author}{G\"{u}neyisi\xfnm[ E.M.]}, \bibinfo{author}{Altay\xfnm[ G.]}.
	\newblock \bibinfo{title}{{Seismic fragility assessment of effectiveness of
			viscous dampers in R/C buildings under scenario earthquakes}}.
	\newblock \bibinfo{journal}{Structural Safety}
	\bibinfo{year}{2008};\bibinfo{volume}{30}(\bibinfo{number}{5}):\bibinfo{pages}{461--480}.
	\bibitem{Seyedi2010}
	\bibinfo{author}{Seyedi\xfnm[ D.M.]}, \bibinfo{author}{Gehl\xfnm[ P.]},
	\bibinfo{author}{Douglas\xfnm[ J.]}, \bibinfo{author}{Davenne\xfnm[ L.]},
	\bibinfo{author}{Mezher\xfnm[ N.]}, \bibinfo{author}{Ghavamian\xfnm[ S.]}.
	\newblock \bibinfo{title}{{Development of seismic fragility surfaces for
			reinforced concrete buildings by means of nonlinear time-history analysis}}.
	\newblock \bibinfo{journal}{Earthquake Eng Struct Dyn}
	\bibinfo{year}{2010};\bibinfo{volume}{39}(\bibinfo{number}{1}):\bibinfo{pages}{91--108}.
	\bibitem{Gardoni2002a}
	\bibinfo{author}{Gardoni\xfnm[ P.]}, \bibinfo{author}{{Der Kiureghian}\xfnm[
		A.]}, \bibinfo{author}{Mosalam\xfnm[ K.M.]}.
	\newblock \bibinfo{title}{{Probabilistic capacity models and fragility
			estimates for reinforced concrete columns based on experimental
			observations}}.
	\newblock \bibinfo{journal}{J Eng Mech}
	\bibinfo{year}{2002};\bibinfo{volume}{128}(\bibinfo{number}{10}):\bibinfo{pages}{1024--1038}.
	\bibitem{Ghosh2010}
	\bibinfo{author}{Ghosh\xfnm[ J.]}, \bibinfo{author}{Padgett\xfnm[ J.E.]}.
	\newblock \bibinfo{title}{{Aging considerations in the development of
			time-dependent seismic fragility curves}}.
	\newblock \bibinfo{journal}{J Struct Eng}
	\bibinfo{year}{2010};\bibinfo{volume}{136}(\bibinfo{number}{12}):\bibinfo{pages}{1497--1511}.
	\bibitem{Argyroudis2012}
	\bibinfo{author}{Argyroudis\xfnm[ S.]}, \bibinfo{author}{Pitilakis\xfnm[ K.]}.
	\newblock \bibinfo{title}{{Seismic fragility curves of shallow tunnels in
			alluvial deposits}}.
	\newblock \bibinfo{journal}{Soil Dyn Earthq Eng}
	\bibinfo{year}{2012};\bibinfo{volume}{35}:\bibinfo{pages}{1--12}.
	\bibitem{Chiou2011}
	\bibinfo{author}{Chiou\xfnm[ J.]}, \bibinfo{author}{Chiang\xfnm[ C.]},
	\bibinfo{author}{Yang\xfnm[ H.]}, \bibinfo{author}{Hsu\xfnm[ S.]}.
	\newblock \bibinfo{title}{{Developing fragility curves for a pile-supported
			wharf}}.
	\newblock \bibinfo{journal}{Soil Dyn Earthq Eng}
	\bibinfo{year}{2011};\bibinfo{volume}{31}:\bibinfo{pages}{830--840}.
	\bibitem{Quilligan2012}
	\bibinfo{author}{Quilligan\xfnm[ A.]}, \bibinfo{author}{{O Connor}\xfnm[ A.]},
	\bibinfo{author}{Pakrashi\xfnm[ V.]}.
	\newblock \bibinfo{title}{{Fragility analysis of steel and concrete wind
			turbine towers}}.
	\newblock \bibinfo{journal}{Engineering Structures}
	\bibinfo{year}{2012};\bibinfo{volume}{36}:\bibinfo{pages}{270--282}.
	\bibitem{Borgonovo2013}
	\bibinfo{author}{Borgonovo\xfnm[ E.]}, \bibinfo{author}{Zentner\xfnm[ I.]},
	\bibinfo{author}{Pellegri\xfnm[ A.]}, \bibinfo{author}{Tarantola\xfnm[ S.]},
	\bibinfo{author}{de~Rocquigny\xfnm[ E.]}.
	\newblock \bibinfo{title}{{On the importance of uncertain factors in seismic
			fragility assessment}}.
	\newblock \bibinfo{journal}{Reliab Eng Sys Safety}
	\bibinfo{year}{2013};\bibinfo{volume}{109}(\bibinfo{number}{0}):\bibinfo{pages}{66--76}.
	\bibitem{Karantoni2014}
	\bibinfo{author}{Karantoni\xfnm[ F.]}, \bibinfo{author}{Tsionis\xfnm[ G.]},
	\bibinfo{author}{Lyrantzaki\xfnm[ F.]}, \bibinfo{author}{Fardis\xfnm[ M. N.]}.
	\newblock \bibinfo{title}{{Seismic fragility of regular masonry buildings for in-plane and out-of-plane failure}}.
	\newblock \bibinfo{journal}{Earthq Struct}
	\bibinfo{year}{2014};\bibinfo{volume}{6}(\bibinfo{number}{6}):\bibinfo{pages}{689--713}.
	\bibitem{Rossetto2005}
	\bibinfo{author}{Rossetto\xfnm[ T.]}, \bibinfo{author}{Elnashai\xfnm[ A.]}.
	\newblock \bibinfo{title}{{A new analytical procedure for the derivation of
			displacement-based vulnerability curves for populations of RC structures}}.
	\newblock \bibinfo{journal}{Engineering Structures}
	\bibinfo{year}{2005};\bibinfo{volume}{27}(\bibinfo{number}{3}):\bibinfo{pages}{397--409}.
	\bibitem{Shinozuka2000b}
	\bibinfo{author}{Shinozuka\xfnm[ M.]}, \bibinfo{author}{Feng\xfnm[ M.]},
	\bibinfo{author}{Lee\xfnm[ J.]}, \bibinfo{author}{Naganuma\xfnm[ T.]}.
	\newblock \bibinfo{title}{{Statistical analysis of fragility curves}}.
	\newblock \bibinfo{journal}{J Eng Mech}
	\bibinfo{year}{2000};\bibinfo{volume}{126}(\bibinfo{number}{12}):\bibinfo{pages}{1224--1231}.
	\bibitem{Ellingwood2001}
	\bibinfo{author}{Ellingwood\xfnm[ B.R.]}.
	\newblock \bibinfo{title}{{Earthquake risk assessment of building structures}}.
	\newblock \bibinfo{journal}{Reliab Eng Sys Safety}
	\bibinfo{year}{2001};\bibinfo{volume}{74}(\bibinfo{number}{3}):\bibinfo{pages}{251--262}.
	\bibitem{Zentner2010a}
	\bibinfo{author}{Zentner\xfnm[ I.]}.
	\newblock \bibinfo{title}{{Numerical computation of fragility curves for NPP
			equipment}}.
	\newblock \bibinfo{journal}{Nuc Eng Des}
	\bibinfo{year}{2010};\bibinfo{volume}{240}:\bibinfo{pages}{1614--1621}.
	\bibitem{Gencturk2008}
	\bibinfo{author}{Gencturk\xfnm[ B.]}, \bibinfo{author}{Elnashai\xfnm[ A.]},
	\bibinfo{author}{Song\xfnm[ J.]}.
	\newblock \bibinfo{title}{{Fragility relationships for populations of woodframe
			structures based on inelastic response}}.
	\newblock \bibinfo{journal}{J Earthq Eng}
	\bibinfo{year}{2008};\bibinfo{volume}{12}:\bibinfo{pages}{119--128}.
	\bibitem{Jeong2012}
	\bibinfo{author}{Jeong\xfnm[ S.H.]}, \bibinfo{author}{Mwafy\xfnm[ A.M.]},
	\bibinfo{author}{Elnashai\xfnm[ A.S.]}.
	\newblock \bibinfo{title}{{Probabilistic seismic performance assessment of
			code-compliant multi-story RC buildings}}.
	\newblock \bibinfo{journal}{Engineering Structures}
	\bibinfo{year}{2012};\bibinfo{volume}{34}:\bibinfo{pages}{527--537}.
	\bibitem{Banerjee2008}
	\bibinfo{author}{Banerjee\xfnm[ S.]}, \bibinfo{author}{Shinozuka\xfnm[ M.]}.
	\newblock \bibinfo{title}{{Mechanistic quantification of RC bridge damage
			states under earthquake through fragility analysis}}.
	\newblock \bibinfo{journal}{Prob Eng Mech}
	\bibinfo{year}{2008};\bibinfo{volume}{23}(\bibinfo{number}{1}):\bibinfo{pages}{12--22}.
	\bibitem{Karamlou2015}
	\bibinfo{author}{Karamlou\xfnm[ A.]}, \bibinfo{author}{Bocchini\xfnm[ P.]}.
	\newblock \bibinfo{title}{{Computation of bridge seismic fragility by
			large-scale simulation for probabilistic resilience analysis}}.
	\newblock \bibinfo{journal}{Earthquake Eng Struct Dyn}
	\bibinfo{year}{2015};\bibinfo{volume}{44}(\bibinfo{number}{12}):\bibinfo{pages}{1959--1978}.
	\bibitem{MaiEurodyn2014}
	\bibinfo{author}{Mai\xfnm[ C.V.]}, \bibinfo{author}{Sudret\xfnm[ B.]},
	\bibinfo{author}{Mackie\xfnm[ K.]}, \bibinfo{author}{Stojadinovic\xfnm[ B.]},
	\bibinfo{author}{Konakli\xfnm[ K.]}.
	\newblock \bibinfo{title}{{Non parametric fragility curves for bridges using
			recorded ground motions}}.
	\newblock In: \bibinfo{editor}{Cunha\xfnm[ A.]}, \bibinfo{editor}{Caetano\xfnm[
		E.]}, \bibinfo{editor}{Ribeiro\xfnm[ P.]}, \bibinfo{editor}{M\"{u}ller\xfnm[
		G.]}, editors. \bibinfo{booktitle}{IX International Conference on Structural
		Dynamics, Porto, Portugal}. \bibinfo{year}{2014}, p.
	\bibinfo{pages}{2831--2838}.
	\bibitem{Rezaeian2008}
	\bibinfo{author}{Rezaeian\xfnm[ S.]}, \bibinfo{author}{{Der Kiureghian}\xfnm[
		A.]}.
	\newblock \bibinfo{title}{{A stochastic ground motion model with separable
			temporal and spectral nonstationarities}}.
	\newblock \bibinfo{journal}{Earthquake Eng Struct Dyn}
	\bibinfo{year}{2008};\bibinfo{volume}{37}(\bibinfo{number}{13}):\bibinfo{pages}{1565--1584}.
	\bibitem{Choi2004}
	\bibinfo{author}{Choi\xfnm[ E.]}, \bibinfo{author}{DesRoches\xfnm[ R.]},
	\bibinfo{author}{Nielson\xfnm[ B.]}.
	\newblock \bibinfo{title}{{Seismic fragility of typical bridges in moderate
			seismic zones}}.
	\newblock \bibinfo{journal}{Eng Struct}
	\bibinfo{year}{2004};\bibinfo{volume}{26}(\bibinfo{number}{2}):\bibinfo{pages}{187--199}.
	\bibitem{Padgett2008}
	\bibinfo{author}{Padgett\xfnm[ J.E.]}, \bibinfo{author}{DesRoches\xfnm[ R.]}.
	\newblock \bibinfo{title}{{Methodology for the development of analytical
			fragility curves for retrofitted bridges}}.
	\newblock \bibinfo{journal}{Earthquake Eng Struct Dyn}
	\bibinfo{year}{2008};\bibinfo{volume}{37}(\bibinfo{number}{8}):\bibinfo{pages}{1157--1174}.
	\bibitem{Zareian2007}
	\bibinfo{author}{Zareian\xfnm[ F.]}, \bibinfo{author}{Krawinkler\xfnm[ H.]}.
	\newblock \bibinfo{title}{{Assessment of probability of collapse and design for
			collapse safety}}.
	\newblock \bibinfo{journal}{Earthquake Eng Struct Dyn}
	\bibinfo{year}{2007};\bibinfo{volume}{36}(\bibinfo{number}{13}):\bibinfo{pages}{1901--1914}.
	\bibitem{Shome1998}
	\bibinfo{author}{Shome\xfnm[ N.]}, \bibinfo{author}{Cornell\xfnm[ C.A.]},
	\bibinfo{author}{Bazzurro\xfnm[ P.]}, \bibinfo{author}{Carballo\xfnm[ J.E.]}.
	\newblock \bibinfo{title}{{Earthquakes, records, and nonlinear responses}}.
	\newblock \bibinfo{journal}{Earthquake Spectra}
	\bibinfo{year}{1998};\bibinfo{volume}{14}(\bibinfo{number}{3}):\bibinfo{pages}{469--500}.
	\bibitem{Luco2007}
	\bibinfo{author}{Luco\xfnm[ N.]}, \bibinfo{author}{Bazzurro\xfnm[ P.]}.
	\newblock \bibinfo{title}{{Does amplitude scaling of ground motion records
			result in biased nonlinear structural drift responses?}}
	\newblock \bibinfo{journal}{Earthquake Eng Struct Dyn}
	\bibinfo{year}{2007};\bibinfo{volume}{36}(\bibinfo{number}{13}):\bibinfo{pages}{1813--1835}.
	\bibitem{Cimellaro2009}
	\bibinfo{author}{Cimellaro\xfnm[ G.P.]}, \bibinfo{author}{Reinhorn\xfnm[
		A.M.]}, \bibinfo{author}{D'Ambrisi\xfnm[ A.]}, \bibinfo{author}{{De
			Stefano}\xfnm[ M.]}.
	\newblock \bibinfo{title}{{Fragility analysis and seismic record selection}}.
	\newblock \bibinfo{journal}{J Struct Eng}
	\bibinfo{year}{2009};\bibinfo{volume}{137}(\bibinfo{number}{3}):\bibinfo{pages}{379--390}.
	\bibitem{Mehdizadeh2012}
	\bibinfo{author}{Mehdizadeh\xfnm[ M.]}, \bibinfo{author}{Mackie\xfnm[ K.R.]},
	\bibinfo{author}{Nielson\xfnm[ B.G.]}.
	\newblock \bibinfo{title}{{Scaling bias and record selection for fragility
			analysis}}.
	\newblock In: \bibinfo{booktitle}{Proc. 15th World Conf. Earthquake Eng.}
	\bibinfo{year}{2012},.
	\bibitem{Bazzurro1998}
	\bibinfo{author}{Bazzurro\xfnm[ P.]}, \bibinfo{author}{Cornell\xfnm[ C.A.]},
	\bibinfo{author}{Shome\xfnm[ N.]}, \bibinfo{author}{Carballo\xfnm[ J.E.]}.
	\newblock \bibinfo{title}{{Three proposals for characterizing MDOF nonlinear
			seismic response}}.
	\newblock \bibinfo{journal}{J Struct Eng}
	\bibinfo{year}{1998};\bibinfo{volume}{124}(\bibinfo{number}{11}):\bibinfo{pages}{1281--1289}.
	\bibitem{Vamvatsikos2002}
	\bibinfo{author}{Vamvatsikos\xfnm[ D.]}, \bibinfo{author}{Cornell\xfnm[ C.A.]}.
	\newblock \bibinfo{title}{{Incremental dynamic analysis}}.
	\newblock \bibinfo{journal}{Earthquake Eng Struct Dyn}
	\bibinfo{year}{2002};\bibinfo{volume}{31}(\bibinfo{number}{3}):\bibinfo{pages}{491--514}.
	\bibitem{WandJones}
	\bibinfo{author}{Wand\xfnm[ M.]}, \bibinfo{author}{Jones\xfnm[ M.C.]}.
	\newblock \bibinfo{title}{{Kernel smoothing}}.
	\newblock \bibinfo{publisher}{Chapman and Hall}; \bibinfo{year}{1995}.
	\bibitem{DuongThesis2004}
	\bibinfo{author}{Duong\xfnm[ T.]}.
	\newblock \bibinfo{title}{{Bandwidth selectors for multivariate kernel density
			estimation}}.
	\newblock Ph.D. thesis; School of mathematics and Statistics, University of
	Western Australia; \bibinfo{year}{2004}.
	\bibitem{Duong2005}
	\bibinfo{author}{Duong\xfnm[ T.]}, \bibinfo{author}{Hazelton\xfnm[ M.L.]}.
	\newblock \bibinfo{title}{{Cross-validation bandwidth matrices for multivariate
			kernel density estimation}}.
	\newblock \bibinfo{journal}{Scand J Stat}
	\bibinfo{year}{2005};\bibinfo{volume}{32}(\bibinfo{number}{3}):\bibinfo{pages}{485--506}.
	\bibitem{Frankel2000}
	\bibinfo{author}{Frankel\xfnm[ A.D.]}, \bibinfo{author}{Mueller\xfnm[ C.S.]},
	\bibinfo{author}{Barnhard\xfnm[ T.P.]}, \bibinfo{author}{Leyendecker\xfnm[
		E.V.]}, \bibinfo{author}{Wesson\xfnm[ R.L.]}, \bibinfo{author}{Harmsen\xfnm[
		S.C.]}, et~al.
	\newblock \bibinfo{title}{{USGS national seismic hazard maps}}.
	\newblock \bibinfo{journal}{Earthquake spectra}
	\bibinfo{year}{2000};\bibinfo{volume}{16}(\bibinfo{number}{1}):\bibinfo{pages}{1--19}.
	\bibitem{SudretMaiCFM2013}
	\bibinfo{author}{Sudret\xfnm[ B.]}, \bibinfo{author}{Mai\xfnm[ C.V.]}.
	\newblock \bibinfo{title}{{Calcul des courbes de fragilit\'{e} par approches
			non-param\'{e}triques}}.
	\newblock In: \bibinfo{booktitle}{Proc. 21e Congr\`{e}s Fran\c{c}ais de
		M\'{e}canique (CFM21), Bordeaux}. \bibinfo{year}{2013}{\natexlab{a}},.
	\bibitem{Bradley2010}
	\bibinfo{author}{Bradley\xfnm[ B.A.]}, \bibinfo{author}{Lee\xfnm[ D.S.]}.
	\newblock \bibinfo{title}{{Accuracy of approximate methods of uncertainty
			propagation in seismic loss estimation}}.
	\newblock \bibinfo{journal}{Structural Safety}
	\bibinfo{year}{2010};\bibinfo{volume}{32}(\bibinfo{number}{1}):\bibinfo{pages}{13--24}.
	\bibitem{Liel2009}
	\bibinfo{author}{Liel\xfnm[ A.B.]}, \bibinfo{author}{Haselton\xfnm[ C.B.]},
	\bibinfo{author}{Deierlein\xfnm[ G.G.]}, \bibinfo{author}{Baker\xfnm[ J.W.]}.
	\newblock \bibinfo{title}{{Incorporating modeling uncertainties in the
			assessment of seismic collapse risk of buildings}}.
	\newblock \bibinfo{journal}{Structural Safety}
	\bibinfo{year}{2009};\bibinfo{volume}{31}(\bibinfo{number}{2}):\bibinfo{pages}{197--211}.
	\bibitem{Efron1979}
	\bibinfo{author}{Efron\xfnm[ B.]}.
	\newblock \bibinfo{title}{{Bootstrap methods: another look at the Jackknife}}.
	\newblock \bibinfo{journal}{The Annals of Statistics}
	\bibinfo{year}{1979};\bibinfo{volume}{7}(\bibinfo{number}{1}):\bibinfo{pages}{1--26}.
	\bibitem{Kwong2015}
	\bibinfo{author}{Kwong\xfnm[ N.S.]}, \bibinfo{author}{Chopra\xfnm[ A.K.]},
	\bibinfo{author}{McGuire\xfnm[ R.K.]}.
	\newblock \bibinfo{title}{{Evaluation of ground motion selection and
			modification procedures using synthetic ground motions}}.
	\newblock \bibinfo{journal}{Earthquake Eng Struct Dyn}
	\bibinfo{year}{2015};\bibinfo{volume}{44}(\bibinfo{number}{11}):\bibinfo{pages}{1841--1861}.
	\bibitem{Rezaeian2010}
	\bibinfo{author}{Rezaeian\xfnm[ S.]}, \bibinfo{author}{{Der Kiureghian}\xfnm[
		A.]}.
	\newblock \bibinfo{title}{{Simulation of synthetic ground motions for specified
			earthquake and site characteristics}}.
	\newblock \bibinfo{journal}{Earthquake Eng Struct Dyn}
	\bibinfo{year}{2010};\bibinfo{volume}{39}(\bibinfo{number}{10}):\bibinfo{pages}{1155--1180}.
	\bibitem{Vetter2014}
	\bibinfo{author}{Vetter\xfnm[ C.]}, \bibinfo{author}{Taflanidis\xfnm[ A.A.]}.
	\newblock \bibinfo{title}{{Comparison of alternative stochastic ground motion
			models for seismic risk characterization}}.
	\newblock \bibinfo{journal}{Soil Dyn Earthq Eng}
	\bibinfo{year}{2014};\bibinfo{volume}{58}:\bibinfo{pages}{48--65}.
	\bibitem{Boore2003}
	\bibinfo{author}{Boore\xfnm[ D.M.]}.
	\newblock \bibinfo{title}{{Simulation of Ground Motion Using the Stochastic
			Method}}.
	\newblock \bibinfo{journal}{Pure Appl Geophys}
	\bibinfo{year}{2003};\bibinfo{volume}{160}(\bibinfo{number}{3}):\bibinfo{pages}{635--676}.
	\bibitem{EC1}
	\bibinfo{author}{{Eurocode 1}\xfnm[]}.
	\newblock \bibinfo{title}{{Actions on structures - Part 1-1: general actions -
			densities, self-weight, imposed loads for buildings}}.
	\newblock \bibinfo{year}{2004}.
	\bibitem{OpenSees}
	\bibinfo{author}{{Pacific Earthquake Engineering and Research Center}\xfnm[]}.
	\newblock \bibinfo{title}{{OpenSees: The Open System for Earthquake Engineering
			Simulation}}.
	\newblock \bibinfo{year}{2004}.
	\bibitem{EC3}
	\bibinfo{author}{{Eurocode 3}\xfnm[]}.
	\newblock \bibinfo{title}{{Design of steel structures - Part 1-1: General rules
			and rules for buildings}}.
	\newblock \bibinfo{year}{2005}.
	\bibitem{jcss}
	\bibinfo{author}{{Joint Committee on Structural Safety}\xfnm[]}.
	\newblock \bibinfo{title}{{Probabilistic Model Code - Part 3 : Resistance
			Models}}.
	\newblock \bibinfo{year}{2001}.
	\bibitem{Deierlein2010}
	\bibinfo{author}{Deierlein\xfnm[ G.G.]}, \bibinfo{author}{Reinhorn\xfnm[
		A.M.]}, \bibinfo{author}{Willford\xfnm[ M.R.]}.
	\newblock \bibinfo{title}{{Nonlinear structural analysis for seismic design}}.
	\newblock \bibinfo{journal}{NEHRP Seismic Design Technical Brief No}
	\bibinfo{year}{2010};\bibinfo{volume}{4}.
	\bibitem{Mackie2004}
	\bibinfo{author}{Mackie\xfnm[ K.]}, \bibinfo{author}{Stojadinovic\xfnm[ B.]}.
	\newblock \bibinfo{title}{{Improving probabilistic seismic demand models
			through refined intensity measures}}.
	\newblock In: \bibinfo{booktitle}{Proc. 13th World Conf. Earthquake Eng.}
	\bibinfo{publisher}{Int. Assoc. for Earthquake Eng. Japan};
	\bibinfo{year}{2004},.
	\bibitem{Padgett2008a}
	\bibinfo{author}{Padgett\xfnm[ J.]}, \bibinfo{author}{Nielson\xfnm[ B.]},
	\bibinfo{author}{DesRoches\xfnm[ R.]}.
	\newblock \bibinfo{title}{{Selection of optimal intensity measures in
			probabilistic seismic demand models of highway bridge portfolios}}.
	\newblock \bibinfo{journal}{Earthquake Eng Struct Dyn}
	\bibinfo{year}{2008};\bibinfo{volume}{37}(\bibinfo{number}{5}):\bibinfo{pages}{711--725}.
	\bibitem{Cornell2002}
	\bibinfo{author}{Cornell\xfnm[ C.]}, \bibinfo{author}{Jalayer\xfnm[ F.]},
	\bibinfo{author}{Hamburger\xfnm[ R.]}, \bibinfo{author}{Foutch\xfnm[ D.]}.
	\newblock \bibinfo{title}{{Probabilistic basis for 2000 SAC federal emergency
			management agency steel moment frame guidelines}}.
	\newblock \bibinfo{journal}{J Struct Eng (ASCE)}
	\bibinfo{year}{2002};\bibinfo{volume}{128}(\bibinfo{number}{4}):\bibinfo{pages}{526--533}.
	\bibitem{Lagaros2007}
	\bibinfo{author}{Lagaros\xfnm[ N.D.]}, \bibinfo{author}{Fragiadakis\xfnm[ M.]}.
	\newblock \bibinfo{title}{{Fragility assessment of steel frames using neural
			networks}}.
	\newblock \bibinfo{journal}{Earthquake Spectra}
	\bibinfo{year}{2007};\bibinfo{volume}{23}(\bibinfo{number}{4}):\bibinfo{pages}{735--752}.
	\bibitem{FEMA2000}
	\bibinfo{author}{{Federal Emergency Management Agency, Washington, DC}\xfnm[]}.
	\newblock \bibinfo{title}{{Commentary for the seismic rehabilitation of
			buildings}}; \bibinfo{year}{2000}.
	\bibitem{EC8eng}
	\bibinfo{author}{{Eurocode 8}\xfnm[]}.
	\newblock \bibinfo{title}{{Design of structures for earthquake resistance -
			Part 1: General rules, seismic actions and rules for buildings}};
	\bibinfo{year}{2004}.
	\bibitem{Mackie2003}
	\bibinfo{author}{Mackie\xfnm[ K.]}, \bibinfo{author}{Stojadinovic\xfnm[ B.]}.
	\newblock \bibinfo{title}{{Seismic demands for performance-based design of
			bridges}}.
	\newblock \bibinfo{type}{Tech. Rep.}; Pacific Earthquake Engineering Research
	Center; \bibinfo{year}{2003}.
	\bibitem{Ramamoorthy2006}
	\bibinfo{author}{Ramamoorthy\xfnm[ S.K.]}, \bibinfo{author}{Gardoni\xfnm[ P.]},
	\bibinfo{author}{Bracci\xfnm[ J.]}.
	\newblock \bibinfo{title}{{Probabilistic demand models and fragility curves for
			reinforced concrete frames}}.
	\newblock \bibinfo{journal}{J Struct Eng}
	\bibinfo{year}{2006};\bibinfo{volume}{132}(\bibinfo{number}{10}):\bibinfo{pages}{1563--1572}.
	\bibitem{Bai2011}
	\bibinfo{author}{Bai\xfnm[ J.W.]}, \bibinfo{author}{Gardoni\xfnm[ P.]},
	\bibinfo{author}{Hueste\xfnm[ M.D.]}.
	\newblock \bibinfo{title}{{Story-specific demand models and seismic fragility
			estimates for multi-story buildings}}.
	\newblock \bibinfo{journal}{Structural Safety}
	\bibinfo{year}{2011};\bibinfo{volume}{33}(\bibinfo{number}{1}):\bibinfo{pages}{96--107}.
	\bibitem{Muggeo2003}
	\bibinfo{author}{Muggeo\xfnm[ V.M.R.]}.
	\newblock \bibinfo{title}{{Estimating regression models with unknown
			break-points}}.
	\newblock \bibinfo{journal}{Statistics in Medicine}
	\bibinfo{year}{2003};\bibinfo{volume}{22}:\bibinfo{pages}{3055--3071}.
	\bibitem{Duongks2007}
	\bibinfo{author}{Duong\xfnm[ T.]}.
	\newblock \bibinfo{title}{{ks: kernel density estimation and kernel
			discriminant analysis for multivariate data in R}}.
	\newblock \bibinfo{journal}{J Stat Softw}
	\bibinfo{year}{2007};\bibinfo{volume}{21}(\bibinfo{number}{7}):\bibinfo{pages}{1--16}.
	\bibitem{Choun2010}
	\bibinfo{author}{Choun\xfnm[ Y.S.]}, \bibinfo{author}{Elnashai\xfnm[ A.S.]}.
	\newblock \bibinfo{title}{{A simplified framework for probabilistic earthquake
			loss estimation}}.
	\newblock \bibinfo{journal}{Prob Eng Mech}
	\bibinfo{year}{2010};\bibinfo{volume}{25}(\bibinfo{number}{4}):\bibinfo{pages}{355--364}.
	\bibitem{Marsh2013}
	\bibinfo{author}{Marsh\xfnm[ M.L.]}, \bibinfo{author}{Stringer\xfnm[ S.J.]}.
	\newblock \bibinfo{title}{{Performance-based seismic bridge design, a synthesis
			of highway practice}}; vol. \bibinfo{volume}{440}.
	\newblock \bibinfo{publisher}{Transportation Research Board, Washington D.C.};
	\bibinfo{year}{2013}.
	\bibitem{Lu2005}
	\bibinfo{author}{Lu\xfnm[ Y.]}, \bibinfo{author}{Gu\xfnm[ X.]},
	\bibinfo{author}{Guan\xfnm[ J.]}.
	\newblock \bibinfo{title}{{Probabilistic drift limits and performance
			evaluation of reinforced concrete columns}}.
	\newblock \bibinfo{journal}{J Struct Eng}
	\bibinfo{year}{2005};\bibinfo{volume}{131}(\bibinfo{number}{6}):\bibinfo{pages}{966--978}.
	\bibitem{Jankovic2004}
	\bibinfo{author}{Jankovic\xfnm[ S.]}, \bibinfo{author}{Stojadinovic\xfnm[ B.]}.
	\newblock \bibinfo{title}{{Probabilistic performance based seismic demand model
			for R/C frame buildings}}.
	\newblock In: \bibinfo{booktitle}{Proc. 13th World Conf. Earthquake Eng.}
	\bibinfo{year}{2004},.
	\bibitem{Jalayer2009}
	\bibinfo{author}{Jalayer\xfnm[ F.]}, \bibinfo{author}{Cornell\xfnm[ C.A.]}.
	\newblock \bibinfo{title}{{Alternative non-linear demand estimation methods for
			probability-based seismic assessments}}.
	\newblock \bibinfo{journal}{Earthquake Eng Struct Dyn}
	\bibinfo{year}{2009};\bibinfo{volume}{38}(\bibinfo{number}{8}):\bibinfo{pages}{951--972}.
	\bibitem{Baker2007}
	\bibinfo{author}{Baker\xfnm[ J.W.]}.
	\newblock \bibinfo{title}{{Probabilistic structural response assessment using
			vector-valued intensity measures}}.
	\newblock \bibinfo{journal}{Earthquake Eng Struct Dyn}
	\bibinfo{year}{2007};\bibinfo{volume}{36}(\bibinfo{number}{13}):\bibinfo{pages}{1861--1883}.
	\bibitem{Celik2010}
	\bibinfo{author}{Celik\xfnm[ O.C.]}, \bibinfo{author}{Ellingwood\xfnm[ B.]}.
	\newblock \bibinfo{title}{{Seismic fragilities for non-ductile reinforced
			concrete frames - role of aleatoric and epistemic uncertainties}}.
	\newblock \bibinfo{journal}{Structural Safety}
	\bibinfo{year}{2010};\bibinfo{volume}{32}(\bibinfo{number}{1}):\bibinfo{pages}{1--12}.
	\bibitem{Jalayer2014}
	\bibinfo{author}{Jalayer\xfnm[ F.]}, \bibinfo{author}{{De Risi}\xfnm[ R.]},
	\bibinfo{author}{Manfredi\xfnm[ G.]}.
	\newblock \bibinfo{title}{{Bayesian Cloud Analysis: efficient structural
			fragility assessment using linear regression}}.
	\newblock \bibinfo{journal}{Bulletin of Earthquake Engineering}
	\bibinfo{year}{2014};\bibinfo{volume}{13}(\bibinfo{number}{4}):\bibinfo{pages}{1183--1203}.
	\bibitem{Ghanembook2003}
	\bibinfo{author}{Ghanem\xfnm[ R.]}, \bibinfo{author}{Spanos\xfnm[ P.]}.
	\newblock \bibinfo{title}{{Stochastic Finite Elements : A Spectral Approach}}.
	\newblock \bibinfo{publisher}{Courier Dover Publications};
	\bibinfo{year}{2003}.
	\bibitem{BlatmanJCP2011}
	\bibinfo{author}{Blatman\xfnm[ G.]}, \bibinfo{author}{Sudret\xfnm[ B.]}.
	\newblock \bibinfo{title}{{Adaptive sparse polynomial chaos expansion based on
			Least Angle Regression}}.
	\newblock \bibinfo{journal}{J Comput Phys}
	\bibinfo{year}{2011};\bibinfo{volume}{230}:\bibinfo{pages}{2345--2367}.
	\bibitem{SudretGuyonnet2011}
	\bibinfo{author}{Sudret\xfnm[ B.]}, \bibinfo{author}{Piquard\xfnm[ V.]},
	\bibinfo{author}{Guyonnet\xfnm[ C.]}.
	\newblock \bibinfo{title}{{Use of polynomial chaos expansions to establish
			fragility curves in seismic risk assessment}}.
	\newblock In: \bibinfo{editor}{{G. De Roeck G. Degrande}\xfnm[ G.L.]},
	\bibinfo{editor}{M\"{u}ller\xfnm[ G.]}, editors. \bibinfo{booktitle}{Proc.
		8th Int. Conf. Struct. Dynamics (EURODYN 2011), Leuven, Belgium}.
	\bibinfo{year}{2011},.
	\bibitem{SudretMaiIcossar2013}
	\bibinfo{author}{Sudret\xfnm[ B.]}, \bibinfo{author}{Mai\xfnm[ C.V.]}.
	\newblock \bibinfo{title}{{Computing seismic fragility curves using polynomial
			chaos expansions}}.
	\newblock In: \bibinfo{editor}{Deodatis\xfnm[ G.]}, editor.
	\bibinfo{booktitle}{Proc. 11th Int. Conf. Struct. Safety and Reliability
		(ICOSSAR'2013), New York, USA}. \bibinfo{year}{2013}{\natexlab{b}},.
	
\end{thebibliography}
\end{document}